\documentclass[]{aa} 
\usepackage{graphicx}
\usepackage{txfonts}
\usepackage{hyperref}

\newcommand{\orcid}[1]{\href{https://orcid.org/#1}{\includegraphics[width=10pt]{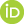}}}
\newcommand{\gaia}{{\em Gaia}}
\begin{document} 

   \title{Gaia data processing. SEAPipe: The source environment analysis pipeline}
   \titlerunning{Gaia data processing. SEAPipe}

   \author{
   D.~L.~Harrison\inst{\ref{inst1},\ref{inst2}}\orcid{0000-0001-8687-6588}
   \and F.~van~Leeuwen\inst{\ref{inst1}}\orcid{0000-0003-1781-4441}
   \and P.~J.~Osborne\inst{\ref{inst1}}\orcid{0000-0003-4482-3538}
   \and P.~W.~Burgess\inst{\ref{inst1}}\orcid{0009-0002-6668-4559}
   \and F.~De~Angeli\inst{\ref{inst1}}\orcid{0000-0003-1879-0488}
   \and D.~W.~Evans\inst{\ref{inst1}}\orcid{0000-0002-6685-5998}
   }

    \institute{Institute of Astronomy, Madingley Road, Cambridge, CB3 0HA, UK. \email{ dlh@ast.cam.ac.uk}\label{inst1}
    \and 
        Kavli Institute for Cosmology, Institute of Astronomy, Madingley Road, Cambridge, CB3 0HA, UK\label{inst2}
    }

   \date{Received ; accepted }

 
  \abstract
   {}
   {To describe two potential options for the Source Environment Analysis pipeline, SEAPipe, for the \gaia{} mission. This pipeline will enable the discovery of sources which are new to \gaia{}, in the sense that they were not found by the on-board detection algorithm. These additional sources (secondaries)  are discoverable in the vicinity of those \gaia{} sources (primaries) that were found by the on-board detection.}
   {The main algorithmic steps required are described; the 2-dimensional image reconstruction of 1-dimensional transit data, the analysis of these images to find the additional sources present, and the determination of the mean positions, proper motions, parallaxes and brightness of these sources.
   Additionally, the Monte Carlo simulations used to characterise the performance of the pipelines are described.
   }
   {The performance of the two options for SEAPipe, the vanilla and image-subtraction versions, are compared. Their selection functions are computed in terms of the magnitude of the secondary sources and their angular separations from their corresponding primary source.
   The completeness and purity of the resultant catalogue of secondary sources as found by each of the pipelines, given the expected magnitude distribution of the primary sources and the magnitude and angular separation distributions of the secondary sources, is also presented.
   The image-subtraction pipeline is shown to out-perform the vanilla pipeline.
   }
   {}


   \maketitle
%

\section{Introduction}
\label{sec:introduction}

On 19 December 2013, the European Space Agency (ESA) launched its \gaia{} satellite \citep{The_Gaia_mission}, which was the start of an ambitious project to measure 
the 3-dimensional spatial and velocity distribution of a billion stars in the Milky Way. \gaia{} started scientific operations in July 2014 and completed the five-year nominal mission on 16 July 2019. As of to date, the spacecraft is in good health and the data collection and processing is still ongoing as an extended mission phase.
The optimisation of the \gaia{} scanning strategy to achieve the best astrometric accuracy leads to its key features. These are the spin rate of 60 ${\rm arcsec\,s}^{-1}$, the maintenance of the spin axis at an angle of $45^{\circ}$ to the Sun, while it slowly precesses around the solar direction, completing a full revolution every 63 days.
The result of the scanning law is that each object would have been observed between 50 to 250 times after the nominal five-year mission, with the ecliptic latitude of the source being the most important factor in determining the coverage.
\gaia{} has two telescopes, and hence two fields-of-view (FoVs), which are projected onto a shared focal plane. 
\gaia{} uses charge-coupled devices, CCDs, its focal plane consisting of 106 of these detectors arranged in seven across-scan (AC) rows and 17 along-scan (AL) strips.
The data used by Source Environment Analysis Pipeline (SEAPipe) originates from the star-mapper, SM, and the astrometric field, AF, CCDs.
A schematic illustration of the focal plane may be found in \cite{The_Gaia_mission}.
For the data rate to be manageable, not all of this CCD data can be transmitted back to Earth.
Regions of the CCDs, known as windows, are assigned by an on-board detection algorithm around the sources it detects.
The size and the level of binning of these windows depends on the CCD and the instantaneous on-board estimate of the magnitude of the detected source; the complete description of window sizes and binning may be found in \cite{2022gdr3.reptE...1D}.
For bright stars these windows are composed of the 2-dimensional pixel data, while for fainter stars this 2-dimensional data is reduced to 1-dimension, by binning this pixel data in the AC (across-scan) direction.
Over the course of the mission the orientation of the focal plane as it passes over each sky-location will vary.
Each source will, therefore, have windows which cross it in different orientations.
These multiple observations may be used to produce a 2-dimensional image of the region surrounding each source from the 1-dimensional window data. 
Indeed in order to achieve the full potential of the \gaia{} mission, this is necessary as nearby undetected companions may otherwise bias the observations of the primary sources (sources found by the on-board detection).
The production of these images allows the detection of any additional (secondary) sources in the vicinity, and allows the necessary corrections to be made to the astrometric and photometric parameters of the primary source.
This is the aim and purpose of SEAPipe.

This paper describes the processing required by SEAPipe to achieve the detection of secondary sources in the vicinity of the primary \gaia{} sources, and to measure the mean positions, proper motions, parallaxes and brightness of these sources, both primary and secondary so as to attain the best possible astrometric and photometric accuracy of the \gaia{} results.
It presents two possible options for SEAPipe, which will be referred to as the vanilla and the image-subtraction pipelines. Both of these pipelines are composed of three main algorithms, image reconstruction, image segregation and image parameter anaylsis.
The image reconstruction forms the 2-dimensional image from which the image segregation finds any additional sources, whose astrometric parameters and brightness are resolved by the image parameter anaylsis. 
The results of SEAPipe processing will be incorporated into the DR4 and DR5 data releases, while results from the image reconstruction and image segregation stages have already been used internally in DR3.
The image reconstruction, image segregation and image parameter anaylsis algorithms are described in Sect.~\ref{sec:seapipe}, as well as the vanilla and the image-subtraction pipelines which connect these steps together to process the \gaia{} data.
The validation and performance assessment of both options for SEAPipe is described in Sect.~\ref{sec:validation}, and the results of these assessments are discussed in Sect.~\ref{sec:discussion}.

\section{SEAPipe}
\label{sec:seapipe}
\begin{figure}
\includegraphics[width=\columnwidth]{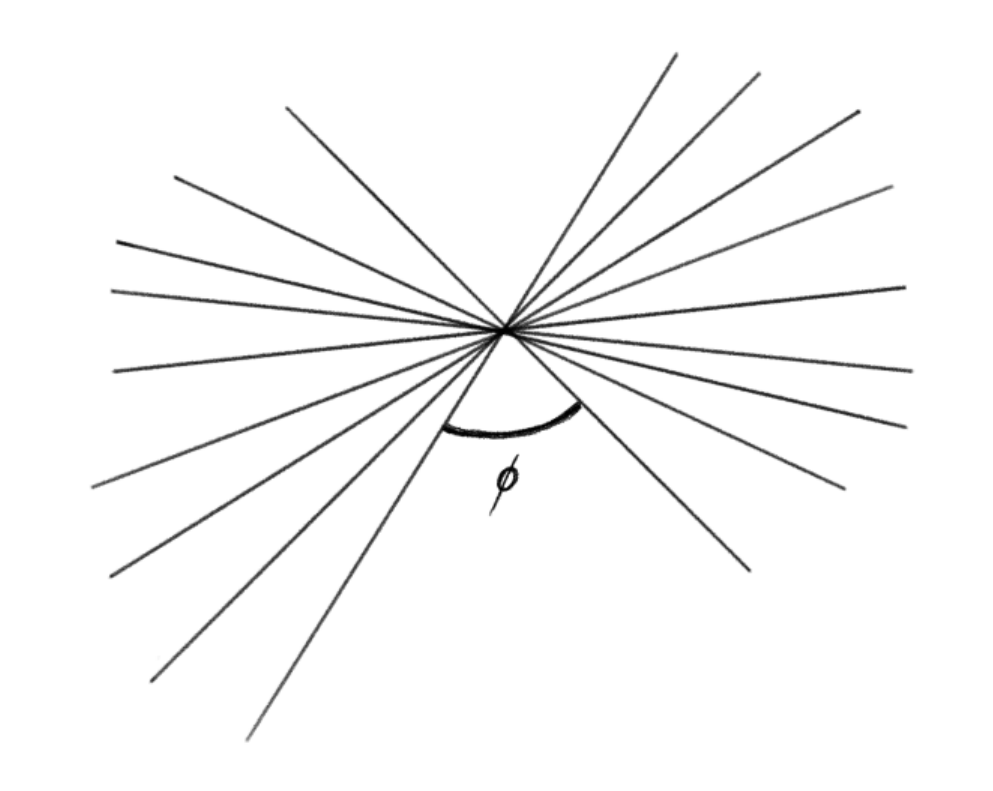}
\caption{Every source will have FoV transits observed in different orientations, as illustrated by this sketch. We define a maximum gap angle, $\phi$,  to be the largest angle between transits with usable data.}
\label{fig:maxGapAngle}
\end{figure}

Before a primary source is analysed by SEAPipe, it must first be assessed whether the source has enough data for processing to proceed. For the image reconstruction step we require at least 10 usable field-of-view (FoV) transits, and a maximum gap angle of $\leq92\deg$, where the maximum gap angle is the largest angle between the scanning directions of the focal place during the transits of the source as illustrated in Fig.~\ref{fig:maxGapAngle}. 
Ideally, this constraint would be $\leq90\deg$, but relaxing this threshold to 92 means all sources, regardless of their position on the sky, are potentially processable by SEAPipe after the five-year nominal mission.
This requirement comes from the image reconstruction which requires observations in different orientations in order to produce a reliable image which is safe to pass to the image segregation step. Without sufficient data the image is likely to contain artefacts which may mistakenly be assessed as additional sources, this is something we want to avoid.
If the maximum gap angle constraint was kept at $\leq90\deg$, sources located in 8.5\,\% of the sky centred on the ecliptic plane would not be processable by SEAPipe.
The resultant drop in the completeness would be dramatic, while the increase in reliability would be minimal to insignificant.

We define a usable FoV transit as a transit on which there is at least a single window that has survived the filtering process, and that this window does not belong to AF1 (Astrometric Field 1). 
The AF1 windows are narrower in the AL (along-scan) direction (see \cite{2022gdr3.reptE...1D}) than the AF2-9 windows and consequently are not used by SEAPipe.
As only three AF windows per transit are used in the image reconstruction, see Sect.~\ref{subsec:imageRecon}, this has an insignificant impact on the number of sources discovered by SEAPipe. 
The filtering process inspects the CCD acquisition level flags and if they are non-nominal, the affected windows are rejected. In addition, windows near charge injections, or which have complex gates are also discarded.
Gates, which effectively reduce the integration time, are activated on-board for bright objects to limit saturation. However, they apply to the full CCD columns containing the gated-window. This means that it is possible for only part of a window to be affected by a gate triggered by other source; these cases are referred to as complex gates \citep{DR1_photometric_data}.
If the LPCs (local plane coordinates), which encode the information on the position of the window samples \citep{LPC}, are not present then these windows also cannot be used. 
While the scanning strategy allows for at least 50 observations of a source over the course of the nominal mission, this may not be the case in practice. The limits on the number of windows which may be read simultaneously from the CCDs limit the number of observations in crowded regions. In addition, some data may be lost when the scanning direction aligns with the Galactic plane if there is not enough bandwidth to transmit all of the data to ground and the on-board storage is exceeded. Due to the prioritisation of data, these effects are more likely to impact fainter sources.
Finally, as part of this data preparation, the windows have their CCD electronic bias and background subtracted, see \cite{GaiaDR1-Pre-processing_and_source_list_creation} for more details on the bias and sources of the background signals.

\subsection{Image Reconstuction}
\label{subsec:imageRecon}

\begin{figure}
\includegraphics[width=\columnwidth]{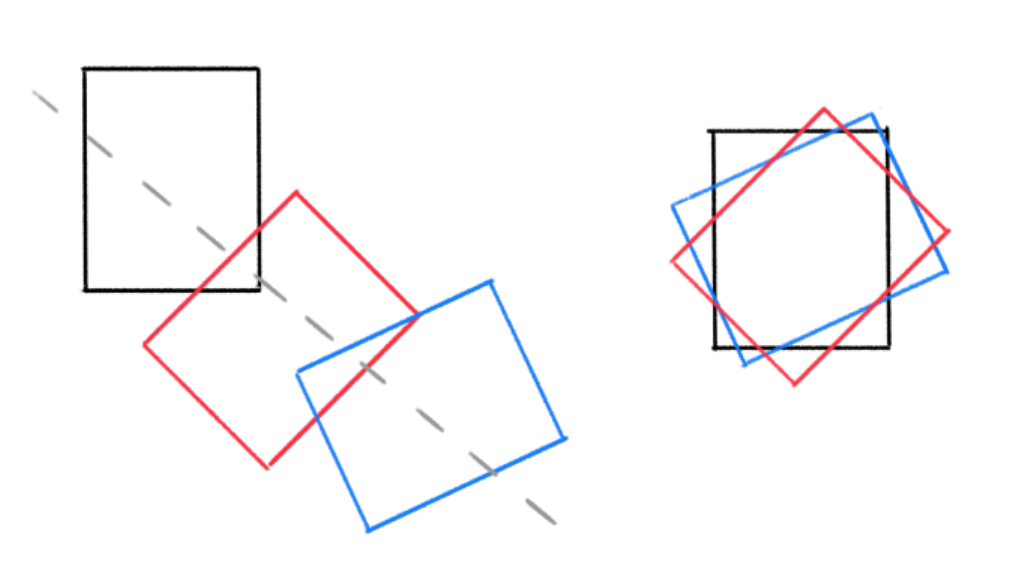}
\caption{Left the path of the primary source on the sky is indicated by the dashed grey line, and windows at three different epochs by the coloured rectangles. In order to obtain an 2-dimensional reconstructed image of this source we need to stack on the position of the source in each window as indicated on the right. Note that the location of the windows for the source on the left is exaggerated for the majority of sources and only reflects the situation for high proper motion sources. However, smaller proper motions could still cause issues, and result in a blurring of the source in the image and a point source could be mistakenly classified as extended. All images are hence formed though stacking on the position of the source in each window.}
\label{fig:stacking}
\end{figure}

The first operation in SEAPipe is image reconstruction, where a 2-dimensional image is formed from the mostly 1-dimensional transit data (AF windows are 1-dimensional for sources with G$>$13~mag, and 2-dimensional otherwise). 
The algorithm used to perform the image reconstruction is described in \cite{harrison11}. 
It essentially stacks the window samples with a weighting system designed to minimise the contamination of the reconstructed image pixels from the 1-dimensional samples dominated by the flux from the primary source in regions of the image which do not contain the primary.
The LPCs provide information on the sky location of the windows; these positions are provided with respect to a reference position ($\alpha_0$, $\delta_0$) which is the barycentric geometric position of the source at the epoch of the specific catalogue used for the generation of the LPCs.
The LPCs also provide the centroid position of the primary source in each window with respect to this reference position.
If the source is moving then stacking the windows based on their offsets from the reference position on the sky would either form an elongated image of the source or no image at all depending on how fast the source is moving.
Instead, the image is formed by stacking on the position of the source in the window as shown in Fig.~\ref{fig:stacking}.
The reconstructed image is now an image co-moving with the source. 
How any secondary sources in the vicinity of the primary, will appear in the image will depend on their relative motion with respect to the primary source.
\begin{figure*}
\includegraphics[width=\textwidth/3]{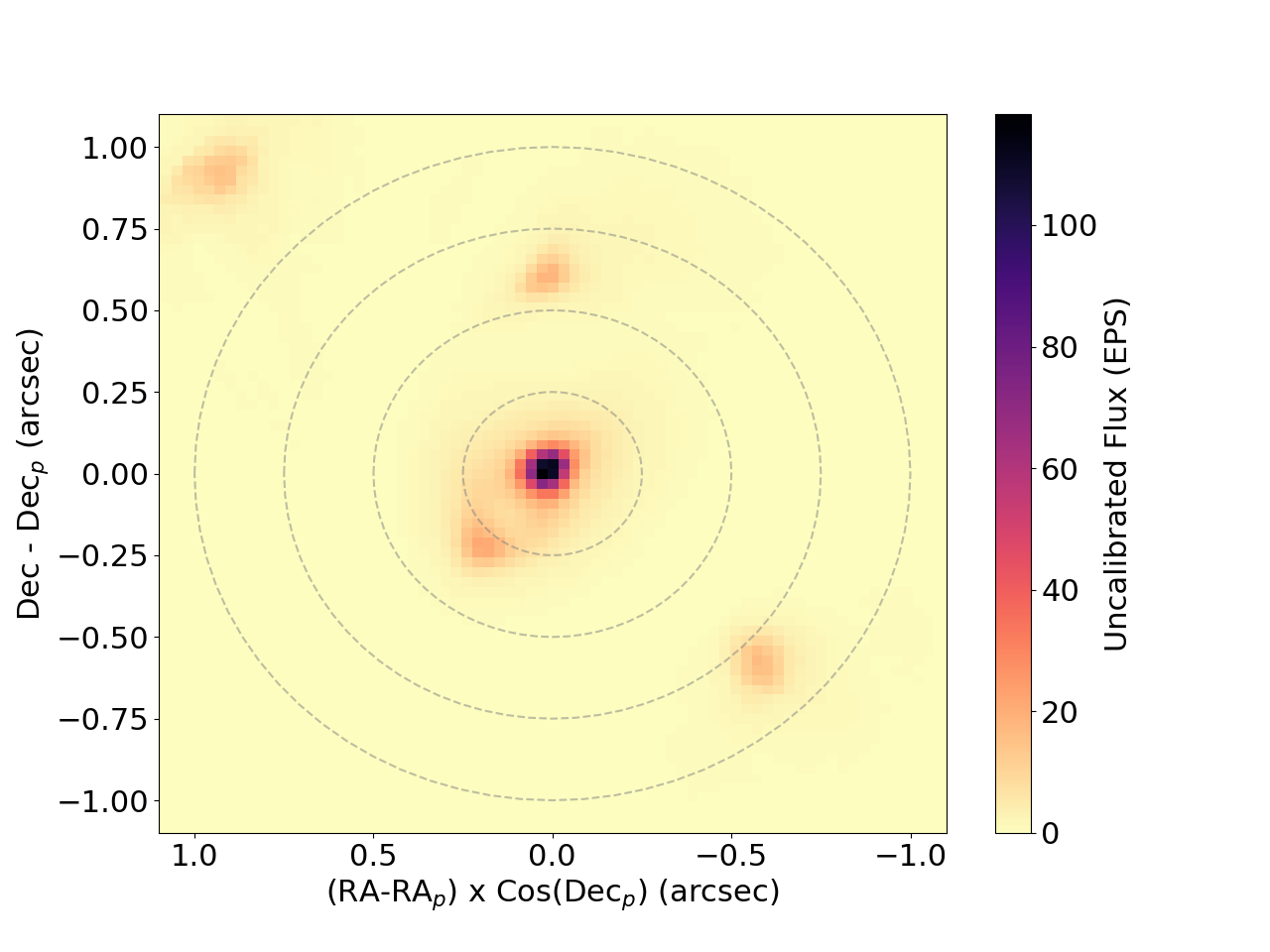}\includegraphics[width=\textwidth/3]{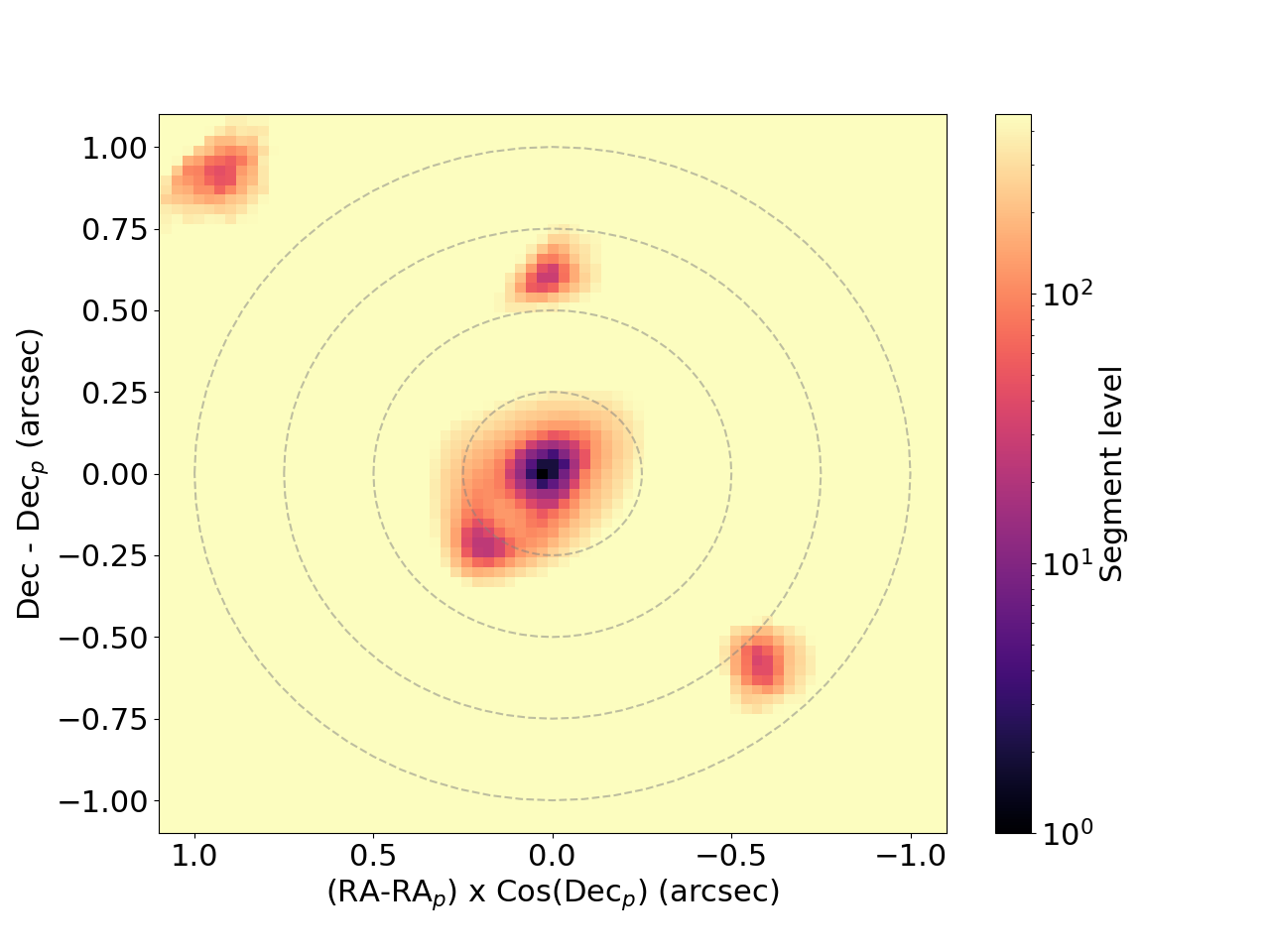} \includegraphics[width=\textwidth/3]{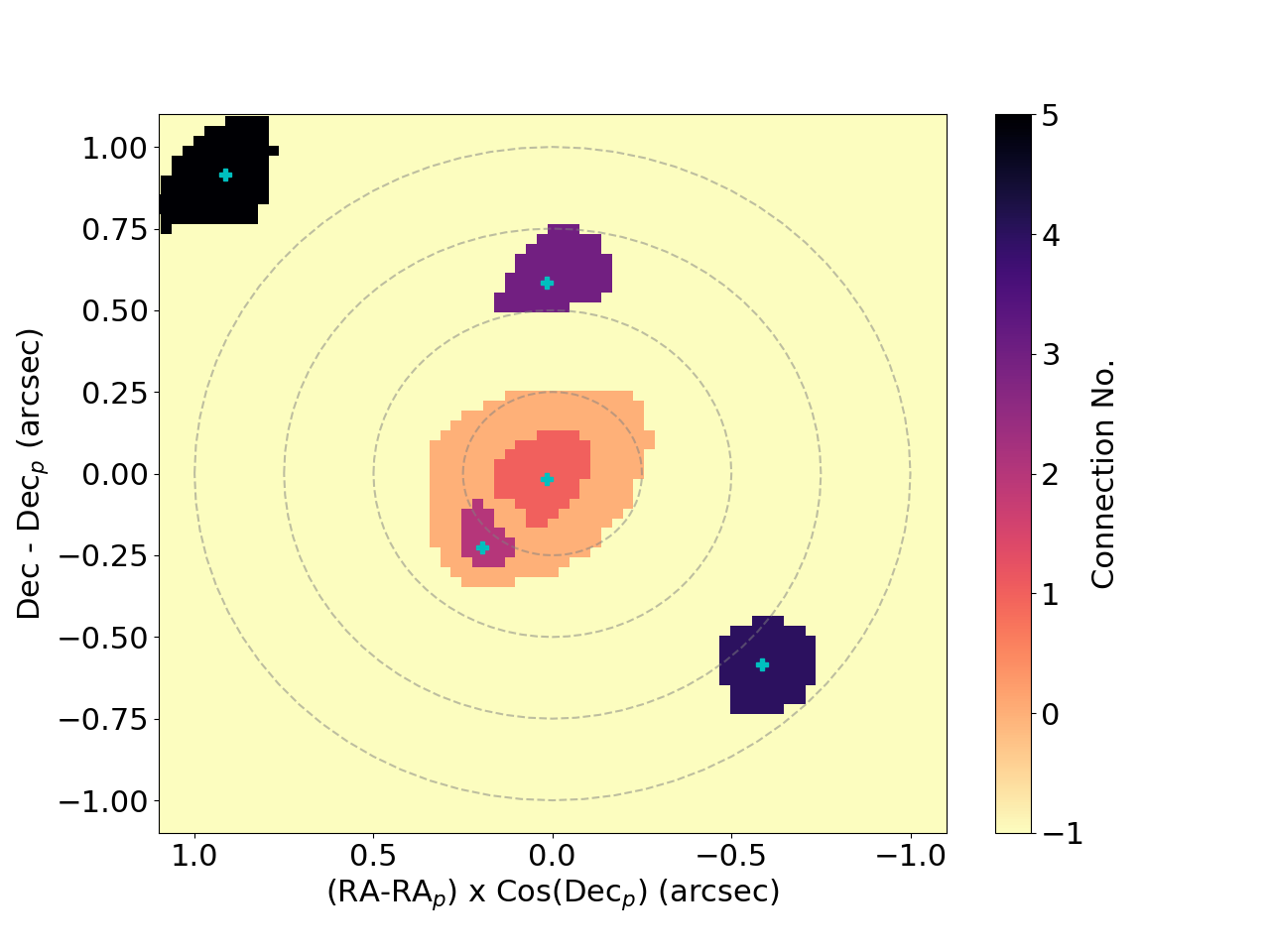}
\caption{Left: reconstructed image of a $G$=14.0 magnitude primary source, located at (${\rm RA}_p$,${\rm Dec}_p$), into whose data four secondary sources, all 2 magnitudes fainter, have been injected. Centre: segmented image, the segment containing the brightest pixel is labelled 1. For clarity, the threshold level at which the segmentation stops has been increased for this plot, and a log-scale is used. Right: segregated image, here the blue crosses indicate the positions returned for the sources found in this image. A connection number of -1 or 0 means this image pixel belongs to the background and not a source.
In all panels, the grey dashed circles indicate angular separations of 0.25, 0.50, 0.75, and 1.0" from the centre of the image (and primary source position).
}
\label{fig:imgseg_example}
\end{figure*}

%
The CPU time taken by the image reconstruction per primary source increases non-linearly with the amount of data used. There is hence a trade-off between the number of windows used in this step and the time taken. Beyond a certain point including additional AF window data from the same transit, does not significantly improve the resultant image (and hence the number of additional sources discoverable by the image segregation step). Including additional data from different orientations is always a good thing.
Monte Carlo simulations, see Sect.~\ref{sec:mcqa}, were used to explore this trade-off. It was determined that using more than three AF windows per transit provides limited returns for CPU resources spent.
The choice of which AF windows to use is set by the desire to spread out the observations along the focal plane.
A priority ordering was chosen in order to best achieve this, so that if a high priority AF window is unusable, a lower priority one will be selected. 
This ensures that three AF windows will always be used, if there are at least three usable AF windows for a given transit.
This is done to avoid all the selected windows from a transit being contaminated by a parasitic observation from the other FoV.
A parasitic observation occurs when a source from the other FoV happens to be projected onto the same location on the AF CCDs, however, as the AC rate is different for the other FoV, this projection only contributes to a few of the AF CCDs along the transit rather than all of them.
The window data containing an example of a parasitic observation may be seen in Fig.~1 of \cite{fast_transients18}.

\subsection{Image Segregation}

The image segregation step runs on the image produced by the image reconstruction.
It is composed of two main parts, the first segments the image into regions of connected pixels of similar flux-levels and the second then groups these regions together to find sources in the image.
An example of the image segmentation and segregation is shown in Fig.~\ref{fig:imgseg_example} together with the reconstructed image.
The sources visible in the image are necessarily above the noise level, hence in retrieving these sources from the image only the intensity (flux values) of the pixels are required.
The aim of the segmentation is to group neighbouring pixels with similar fluxes together, in order to facilitate assigning pixels to sources in the image.

The segmentation proceeds as follows:
\begin{itemize}
    \item {Find the brightest unconnected pixel, $p_i$, if $f_{p_i}>f_{\rm segment}$ then start a new segment and while new pixels connect to this segment:}
    \begin{itemize}
        \item {Get neighbouring pixels (the eight pixels which surround the current pixel, $p_i$).}
        \item {Find the pixel ($p_j$) with the closest flux value to $p_i$.}
        \item {If $f_{p_i}>f_{p_j}$ and $f_{p_j}/f_{p_i}>0.97$ or if $f_{p_j}>f_{p_i}$ and $f_{p_i}/f_{p_j}>0.97$ then $p_j$ joins this segment.}
        \item {If $p_j$ joins the segment:}
        \begin{itemize}
            \item {If pixel $p_j$ is joined to a previous segment, then the segments merge together.}
            \item {Repeat analysis for this joining pixel $p_j$ (so $p_j$ becomes $p_i$ on the next loop).}
        \end{itemize}
    \end{itemize}  
\end{itemize}
This process continues until there are no pixels left to connect or the flux value of the unconnected pixels is less than a user-defined threshold for the segmentation, $f_{\rm segment}$. Hence, at the end of the segmentation process every pixel with a flux above $f_{\rm segment}$ will have been assigned to a segment.\\

The segregation of the segmented image into sources proceeds as follows:
\begin{itemize}
    \item{The first segment is assigned to the first candidate source.}
    \item{The next segment is checked to see whether any of its pixels neighbour (within a distance of two pixels\footnote{Limiting this to direct neighbours (surrounding eight pixels) resulted in fragmented sources in the segregated image.}; so any of the 24 pixels which surround a given pixel in the segment) that of the first candidate source, if they do this segment becomes part of the first candidate source, if they don't this segment become the second candidate source.}
    \item{The subsequent segments are checked to see which candidate sources they neighbour, if any.}
    \begin{itemize}
        \item{If they have no neighbour they become another candidate source.}
        \item{If they neighbour just one candidate source their pixels becomes part of that candidate source.}
        \item{If they have more than one candidate source as a neighbour then this segment is deemed to belong to a higher level background between two sources and is assigned to the background (as are all pixels which didn't get assigned to a segment in the segmentation step), see Fig.~\ref{fig:imgseg_example}.}
    \end{itemize}
    \item{Once all the segments have been assigned to a candidate source or the background, all candidate sources with less than a user-defined number of pixels are rejected and the following parameters are evaluated for the surviving candidates:}
    \begin{itemize}
        \item{The flux, which is estimated from the sum of the values of the pixels.}
        \item{The positions for the candidate sources are estimated by taking the pixel coordinates of the highest flux valued pixel belonging to the candidate source. These pixel coordinates are expressed as offsets ($({\rm RA}-{\rm RA}_p)\times\cos({\rm Dec}_p)$, ${\rm Dec}-{\rm Dec}_p$) from the position of the primary source (${\rm RA}_p$, ${\rm Dec}_p$).}
        \item{The average maximum gap angle for the image pixels which belong to the candidate source. (The maximum gap angle evaluated previously from the transit data will only be valid at the image centre and not across the whole image).}
        \item{The ratio of the sum of the pixels within two different radii from the pixel containing the peak flux value; provided that these pixels are either assigned to the source in question or the background (this value is used to assess whether the primary source should be classified as extended).}
    \end{itemize}
    \item{These candidates are further filtered, keeping those which}
    \begin{itemize}
        \item{Have an average maximum gap angle less than 100 degrees.}
        \item{An estimated $G$ magnitude brighter than 23.0.}
    \end{itemize}
\end{itemize}

By measuring how concentrated the flux is, with the ratio of the flux contained within two different radii, the image segregation can distinguish between point-like and extended sources.
However, this classification of point source or extended source is only performed on the primary sources, as this is the only source in the image we know to be stationary.
Any secondaries in the image may be moving with respect to the primary and so may appear extended when they are in fact not.
In order for the next step of the image parameter analysis to be run, the primary source must be detected and classified as a point source. 
As will be explained further in Sect.~\ref{sec:pipelines}, the prerequisite of a point-like primary is the only requirement in the case of the image-subtraction pipeline, but in the case of the vanilla pipeline, in addition to this, there must be additional surviving sources found.

\subsection{Image Parameter Analysis}

The image parameter analysis (IPA) takes as input the positions and fluxes of the sources found by the image segregation. 
It cannot detect sources itself, it can only refine their positions and fluxes. 
It can reject sources, say for instance if the updated position is moved outside of the range of the image, or the flux is reduced to the noise floor.
Hence, the number of sources reported by the image segregation can only decrease and never increase after the IPA.
The image reconstruction and segregation serve to find and provide initial estimates for the secondary sources, and the IPA uses the window and LPC data to improve upon these initial values. 
The IPA uses a least squares analysis to fit a model of the position and motion of the primary and secondary sources with respect to the catalogue epoch position of the primary source. 

The observed flux, $o_j$, of the $j^{th}$ window sample may be written as:
\begin{equation}
o_j =  \sum_{i}^{N_s} \Biggl( I_{i}\cdot R(w_{i}-w_j,z_{i}-z_j) \Biggr) + \epsilon_j,
\end{equation}  
where $\epsilon_j$ is the flux error, $N_s$ is the number of sources considered, $R$ is the response (PSF) of \gaia{}, $I_{i}$ is the flux of the $i^{th}$ source, and $w_{i}$ ($z_{i}$) is its position in the AL (AC) direction in this window.
The flux of $j^{th}$ window sample may be predicted (modelled) using:
\begin{equation}
\tilde{o}_j =  \sum_{i}^{N_s} \Biggl( \tilde{I}_{i}\cdot R(\tilde{w}_{i}-w_j,\tilde{z}_{i}-z_j) \Biggr) 
\end{equation}  
The differences between the predicted ($\tilde{o}_j$) and observed ($o_j$) window samples can be expressed as a first-order approximation to corrections for $\tilde{w}_{i}$, $\tilde{z}_{i}$ and $\tilde{I}_{i}$:
\begin{equation}
\label{equ:observ}
\begin{split}
o_j - \tilde{o}_j &=\sum_{i}^{N_s} \Biggl(  \mathrm{d}I_{i}\cdot R(\tilde{w}_{i}-w_{j},\tilde{z}_{i}-z_j)  + \tilde{I}_{i}\cdot\frac{\partial R}{\partial w_{i}}\:\mathrm{d}w_{i}  \\
& \quad + \tilde{I}_{i}\cdot\frac{\partial R}{\partial z_{i}}\:\mathrm{d}z_{i} \Biggr) + \epsilon_j,
\end{split}
\end{equation}
The PSF as described in \cite{elsf} is used to generate the expected response, and its derivatives in the AL and AC directions for every window sample.
Using the transformation from the local scan coordinate ($w$, $z$) to the local equatorial coordinates ($a$, $d$) from \cite{LPC}:
\begin{equation}
\begin{split}
a &= w \sin \theta - z \cos \theta  \\
d &= w \cos \theta + z \sin \theta
\end{split}
\label{eqn:ad2wz}
\end{equation}
where $\theta$ is the scan angle, we can convert these corrections to ones in the local equatorial coordinates. 
Note, these local coordinates are on a tangent plane to the sphere with the position of the primary source at the epoch of the catalogue being the tangent (reference) point.

We can then rewrite Eq.~\ref{equ:observ} in terms of corrections to the local equatorial coordinates and extended it to include corrections to the proper motions and parallaxes:
\begin{equation}
\label{equ:obs_full}
\begin{split}
o_j - \tilde{o}_j &= \sum_{i}^{N_s} \Biggl(  \mathrm{d}I_{i}\cdot R(\tilde{w}_{i}-w_{j},\tilde{z}_{i}-z_j)\\
& \quad + \tilde{I}_{i}\biggl[\frac{\partial R}{\partial w_{i}}\sin\theta - \frac{\partial R}{\partial z_{i}}\cos\theta\biggr]\mathrm{d} a_{i}  \\
& \quad + \tilde{I}_{i}\biggl[\frac{\partial R}{\partial w_{i}}\cos\theta + \frac{\partial R}{\partial z_{i}}\sin\theta \biggr]\mathrm{d} d_{i} \\
& \quad +\tilde{I}_{i}\biggl[\frac{\partial R}{\partial w_{i}}\sin\theta - \frac{\partial R}{\partial z_{i}}\cos\theta\biggr] \Delta T \, \mathrm{d} \mu_{a_{i}}  \\
& \quad + \tilde{I}_{i}\biggl[\frac{\partial R}{\partial w_{i}}\cos\theta + \frac{\partial R}{\partial z_{i}}\sin\theta \biggr] \Delta T \, \mathrm{d}\mu_{d_{i}} \\
& \quad + \tilde{I}_{i}\biggl[f_w \cdot \frac{\partial R}{\partial w_{i}}+f_z \cdot \frac{\partial R}{\partial z_{i}} \biggr] \mathrm{d} \varpi_{i}  \Biggr) + \epsilon_j.
\end{split}
\end{equation}
where $\Delta T$ is the time difference from the mission reference time, $f_w$ is the parallax factor in the AL direction, and  $f_z$ is the parallax factor in the AC direction. 
The parallax factor (which is dimensionless) is the parallactic displacement, this is proportional to $\sin \phi$, where $\phi$ is the angle between the source and the Sun.
All of these parameters are provided by the LPC sample data.

It is Eq.~\ref{equ:obs_full} which is solved for in the least squares analysis. This is an iterative procedure. The positions and fluxes found by the image segregation together with the proper motions and parallax of the primary source are used in the first iteration to find the predicted values per source of $\tilde{w}_{i}$, $\tilde{z}_{i}$ and $\tilde{I}_{i}$ in each window.
Each iteration provides corrections to the parameters of flux, position, proper motion and parallax for each source, with iterations continuing until the corrections to all the parameters are less than 10\,\% of the error in the parameters.
In each iteration the values of these parameters from the previous iteration are used to produce the predicted values of $\tilde{w}_{i}$, $\tilde{z}_{i}$ and $\tilde{I}_{i}$ for each window.

The positions (and hence proper motions) are found on the tangent plane and must be converted to those on the sphere using
\begin{equation}
\begin{split}
\tan(\Delta \alpha) &= \frac{\rm LPC_a}{\cos(\delta_0)-{\rm LPC_d}\sin(\delta_0)}\\
\sin(\delta_i) &= \frac{\sin(\delta_0)+{\rm LPC_d}\cos(\delta_0)}{\sqrt{(1+{\rm LPC_a}^2+{\rm LPC_d}^2)}}
\end{split}
\end{equation}
where ${\rm LPC_a}$ and ${\rm LPC_d}$ are the $a$~and~$d$ in Eqs~\ref{eqn:ad2wz}~and~\ref{equ:obs_full}, and $\Delta \alpha = \alpha_i -\alpha_0$, where the position of the reference point is ($\alpha_0$, $\delta_0$) and position found for the source in question (at the epoch of the catalogue) is ($\alpha_i$, $\delta_i$).
However, for the size of the angles involved here 
\begin{equation}
\begin{split}
\Delta \alpha &=\frac{\rm LPC_a}{\cos(\delta_0)}\\
\Delta \delta &= {\rm LPC_d}
\end{split}
\end{equation}
are sufficient, except for positions close to the poles.\\

\subsection{Pipelines}
\label{sec:pipelines}
\begin{figure}
\includegraphics[width=\columnwidth]{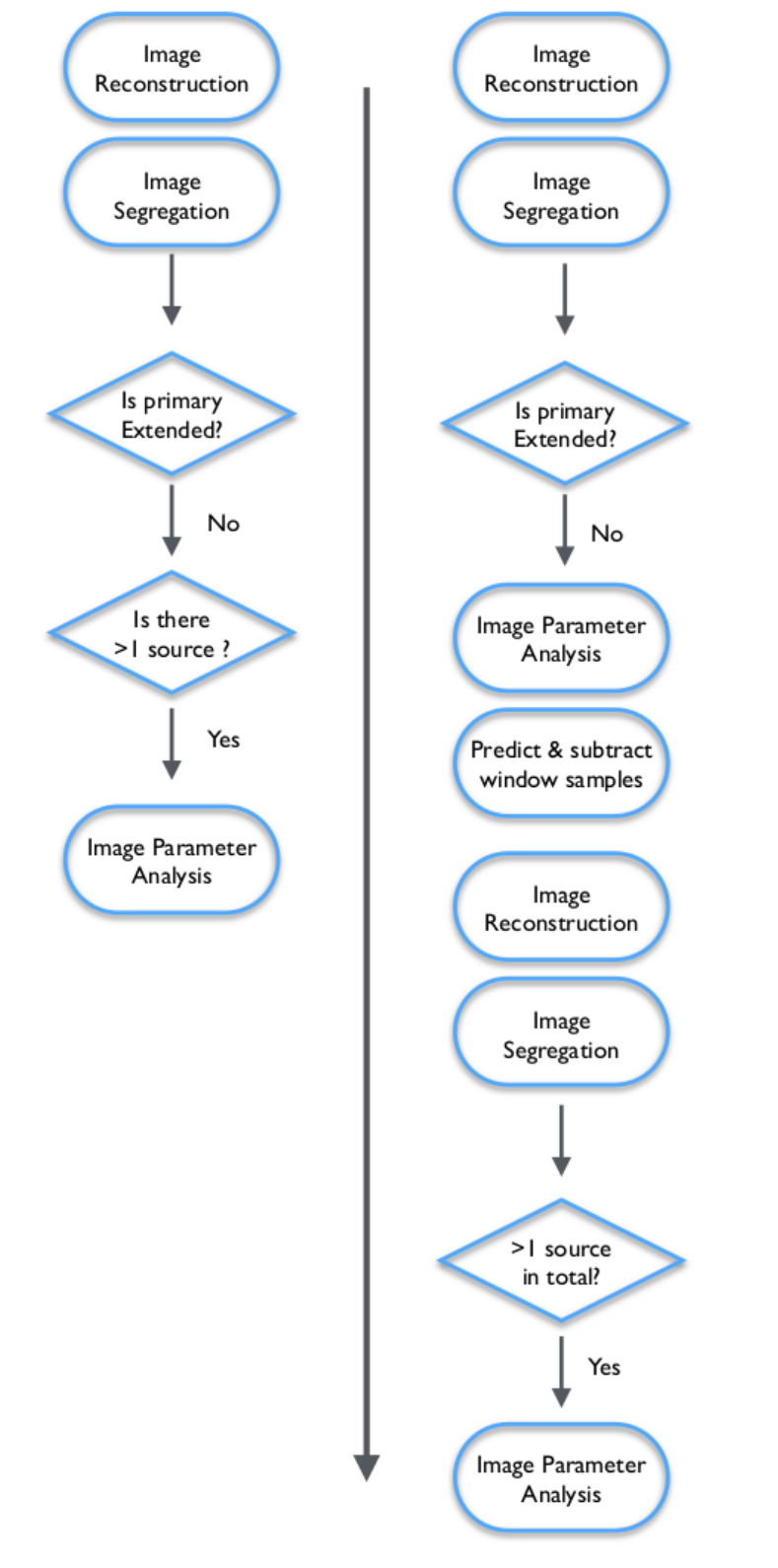}
\caption{Left: the vanilla pipeline. Right: the image-subtraction pipeline. The direction of the arrow indicates the flow of data. The vanilla pipeline is the basic version of SEAPipe, whereas the image-subtraction pipeline seeks to improve the completeness by removing the primary (and any secondaries found in the first pass) from the window samples before repeating the image reconstruction and segregation steps. The final image parameter analysis step includes all the sources found and is performed using the original window sample data.}
\label{fig:dataflow}
\end{figure}
The three main component parts of SEAPipe may be arranged into two alternative pipelines, as shown in Fig.~\ref{fig:dataflow}. 
These are a basic pipeline, known as the vanilla pipeline (on the left of Fig.~\ref{fig:dataflow}), and the image-subtraction pipeline (on the right of Fig.~\ref{fig:dataflow}) which seeks to improve the completeness.

The vanilla pipeline consists of a single pass though the image reconstruction and segregation steps to discover secondary (companion) sources in the vicinity of the primary source. 
If the primary is not extended and there is more than one source present, the image parameter analysis then allows the evaluation of the fluxes and astrometric parameters of all sources found.

The image-subtraction pipeline goes further by taking these fluxes and astrometric parameters and using them to predict the contribution to the window sample data of these sources. These contributions may be subtracted from the window sample data, and the image reconstruction and segregation steps are run again using these residual window sample data.
The aim with these additional steps is to improve the contrast of the resultant images and allow the discovery of fainter sources closer to the primary source.
Hence, this subtraction does not need to be precise and a lower limit on the flux estimates for the sources may be used in order to avoid subtracting too much.
The sources from the first image segregation are kept and merged with the sources detected in the second image segregation, so if any source is re-detected (due to incomplete removal) it will be rejected.
The final image parameter analysis step uses this merged list of sources and is performed using the original window sample data.
The image-subtraction pipeline should thus result in a more complete catalogue of neighbouring sources at the expense of more CPU cycles.
The one concern with the image-subtraction pipeline, however, would be whether the removal of the primary flux from the window sample data could result in artefacts in the resultant reconstructed images which could be mistaken for secondary sources and hence reduce the purity of the catalogue.

The most CPU intensive part of the pipelines is the image parameter analysis. In the case of the vanilla pipeline this is only run when multiple point-like sources have been found.
In the case of the image-subtraction pipeline when only the primary source is found, the first image parameter analysis step may be skipped, using the already known astrometric parameters for the primary source in the prediction and subtraction of window samples step.
Hence the majority of the extra CPU cycles in the image-subtraction pipeline come from the repetition of the image reconstruction and segregation steps.

\section{Validation}
\label{sec:validation}

While it is straightforward to demonstrate successful image reconstruction and segregation (identification of sources in the image) by making comparisons with images from the Hubble Space Telescope archive, this is not sufficient to fully characterise the performance of SEAPipe.

In determining the performance of SEAPipe we would like to characterise its selection function, which describes the probability of detecting a companion (if it exists) at a given angular separation and magnitude difference from the primary source. 
Additionally, we would also like to characterise the completeness and reliability (purity) of the catalogue of companions found by SEAPipe.
The completeness, which may be derived from the selection function if the underlying distribution of companions is known, expresses the fraction of the companions which are detected.
The reliability or purity is the fraction of the companions found which are real sources ($N_{\rm real}/N_{\rm detected}$).
For any algorithm there is a trade-off between completeness and purity, as while lowering the signal-to-noise ratio (S/N) used for detection increases the completeness, this also increases the number of spurious detections, and hence reduces the purity.
Different approaches may have better performances in that they have a greater completeness for a given purity, but the trade-off will still remain.

Ideally, we wish to use the real data to characterise the performance as simulations may not include all of the noise sources present in the real data.
The injection (addition) of fake point sources using the known PSFs of \gaia{} into the real data may be used to evaluate the selection function and completeness of SEAPipe.
If this injection is into the windows belonging to primary sources which are isolated point sources, having no companion within the area to which SEAPipe is sensitive, then this will also allow the reliability of SEAPipe to be tested.
The construction of a list of isolated point sources is described below in Sect.~\ref{sec:isolated}.
The injection of fake sources into the data and the characterisation of the performance of SEAPipe is described in Sect.~\ref{sec:mcqa}.

\subsection{Construction of known isolated source list}
\label{sec:isolated}

The only ancillary data-set which has the required resolution and overlap in observing frequency and wavelength is from the Hubble Space Telescope.
The Hubble source catalogue (HSC) \cite{Whitmore_2016}, was created by combining the data in the Hubble Legacy Archive into a single master catalogue. There have been to date three versions of the catalogue, with each successive version including improvements and the addition of extra data.
The HSC has been cross-matched with itself, providing a table with a list of neighbours within 1 arcsec. This provides a straight-forward way of finding isolated point sources, which is described in Appendix~\ref{appdx:hubble}.
This list of sources however is not sufficient, as it is missing the brighter populations of sources seen by \gaia{} ($G_{\rm mag} \lessapprox 13$).
In order to find a sample of brighter isolated point sources, we used the Hipparcos catalogue \citep{hipparcos1997}.
All Hipparcos sources with anything which indicated a double or multiple system were rejected; in practice this was anything with an entry in the \verb|Multflag| column and anything with an entry in the \verb|Nsys| column.
%
The cross-match table between Hipparcos and \gaia{} was also used \citep{Marrese19}, rejecting anything with number of neighbours$\,>1$ and anything with astrometric parameter$\,\neq5$.
All sources with \verb|duplicated_source| = True, and \verb|astrometric_excess_noise| $\,\neq 0$ were also rejected.
We also applied a Galactic latitude cut of $|b| > 15$ degrees.
A random selection from the remaining sources was then used.
We also include the spectrophotometric standard stars, SPSS, \citep{spss} in our isolated source sample.
The image reconstruction algorithm was run on all sources, and only sources with sufficient data and no obvious companion in the vicinity (the majority of companions found are outside of the  1 arcsec limit) were kept.
This leaves 44\,774 sources, including 114 SPSS ones.

Only primary sources with obvious companions were removed, for the reason that as the companion sources become fainter it becomes a subjective decision as to whether the companion source is real or an artefact. 
Creating a list of primary sources where the data is free from artefacts would artificially boost the performance of SEAPipe in terms of purity for a given completeness and is something we wish to avoid.
Hence, a visual inspection of the images was made to confirm the presence of a companion prior to removal from the list.

\subsection{Monte Carlo analysis}
\label{sec:mcqa}
\begin{figure}
\includegraphics[width=\columnwidth]{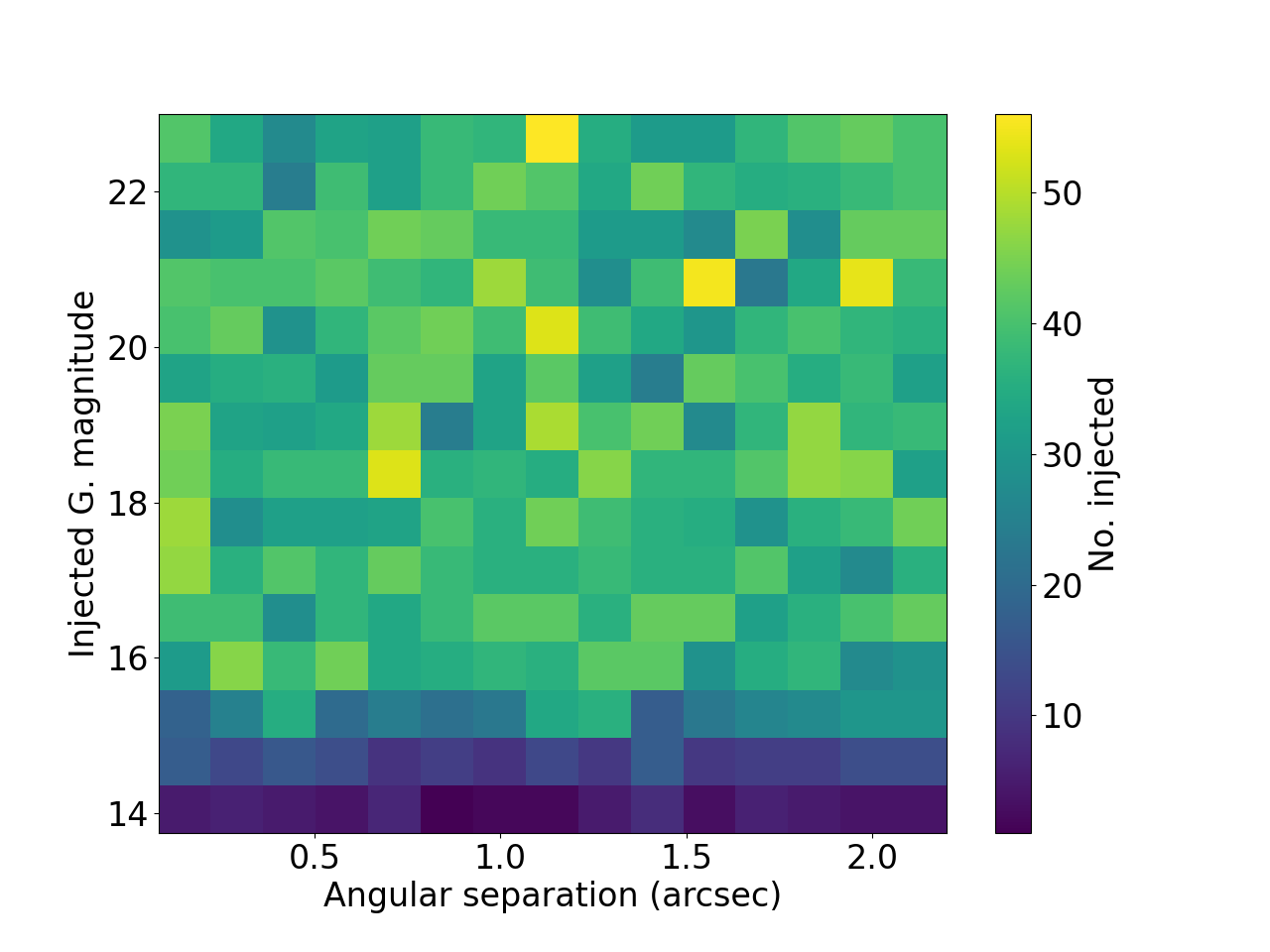}
\caption{The number of injected sources as a function of their magnitude and angular separation from their primary source (in the magnitude range 13.5-15.5). 
The injected secondary sources are drawn from a uniform distribution in magnitude difference and angular separation (from the primary).}
\label{fig:numberInjectedSources}
\end{figure}
%
\begin{figure*}
\includegraphics[width=\textwidth/3]{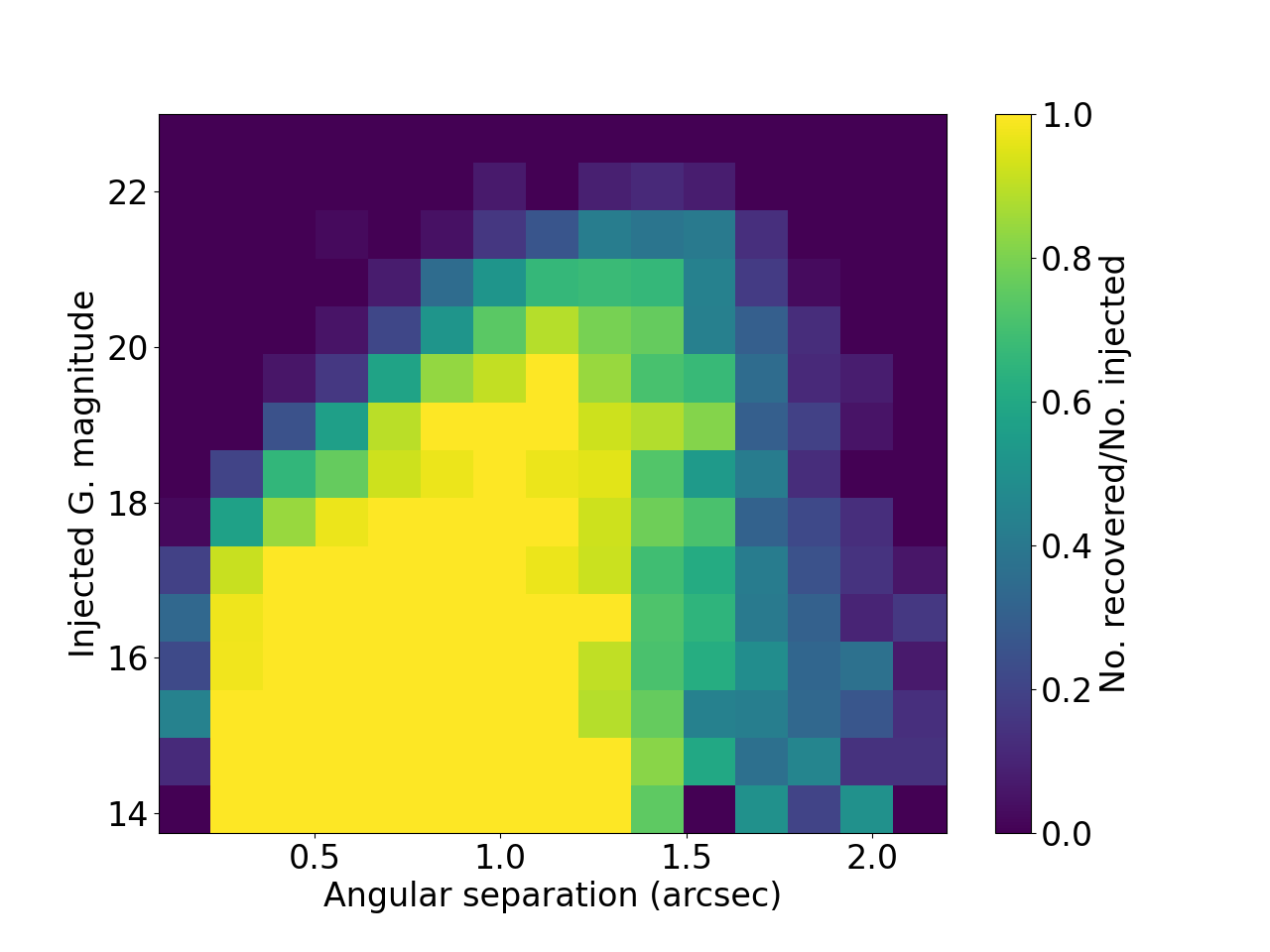}\includegraphics[width=\textwidth/3]{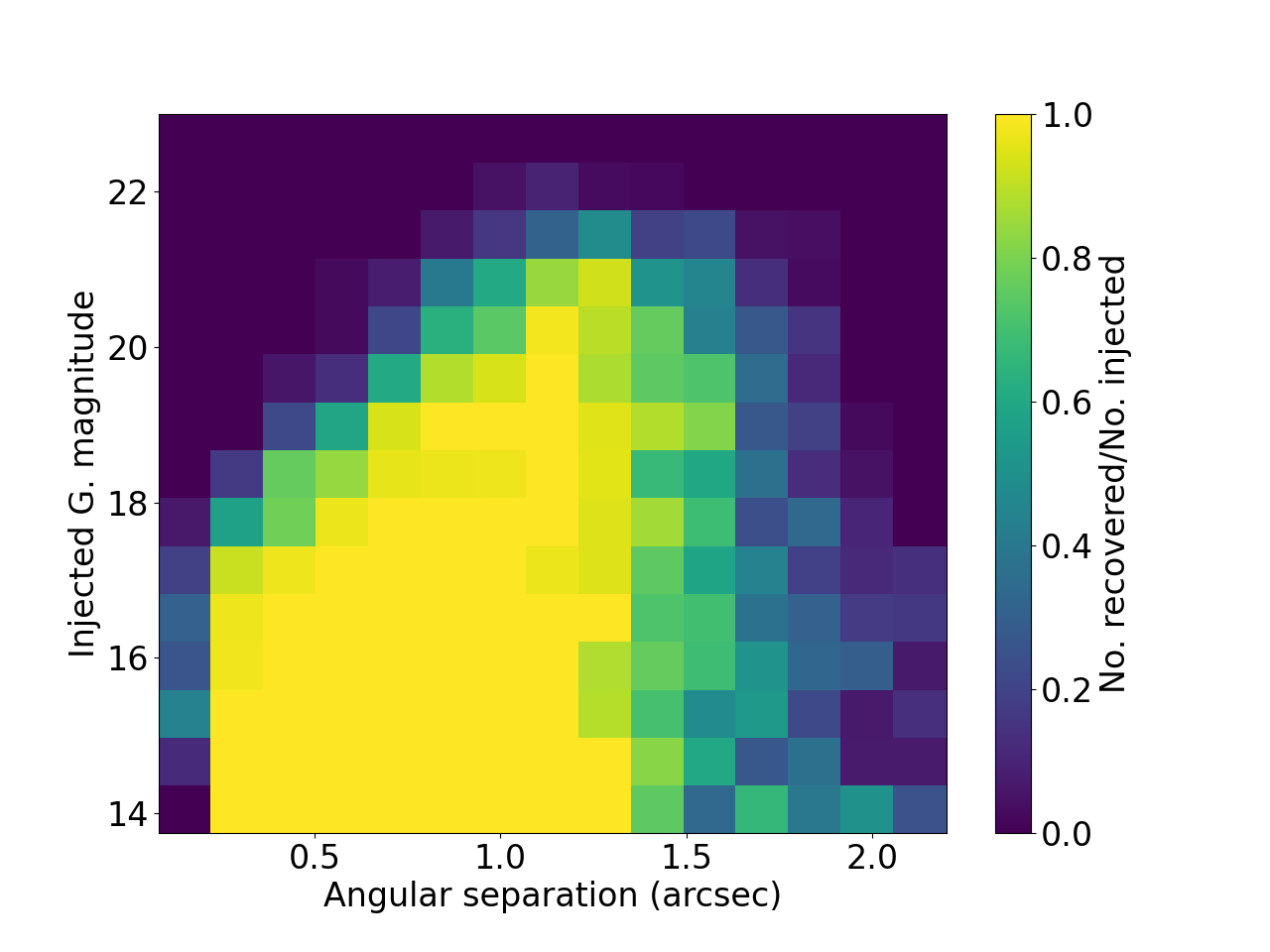}
\includegraphics[width=\textwidth/3]{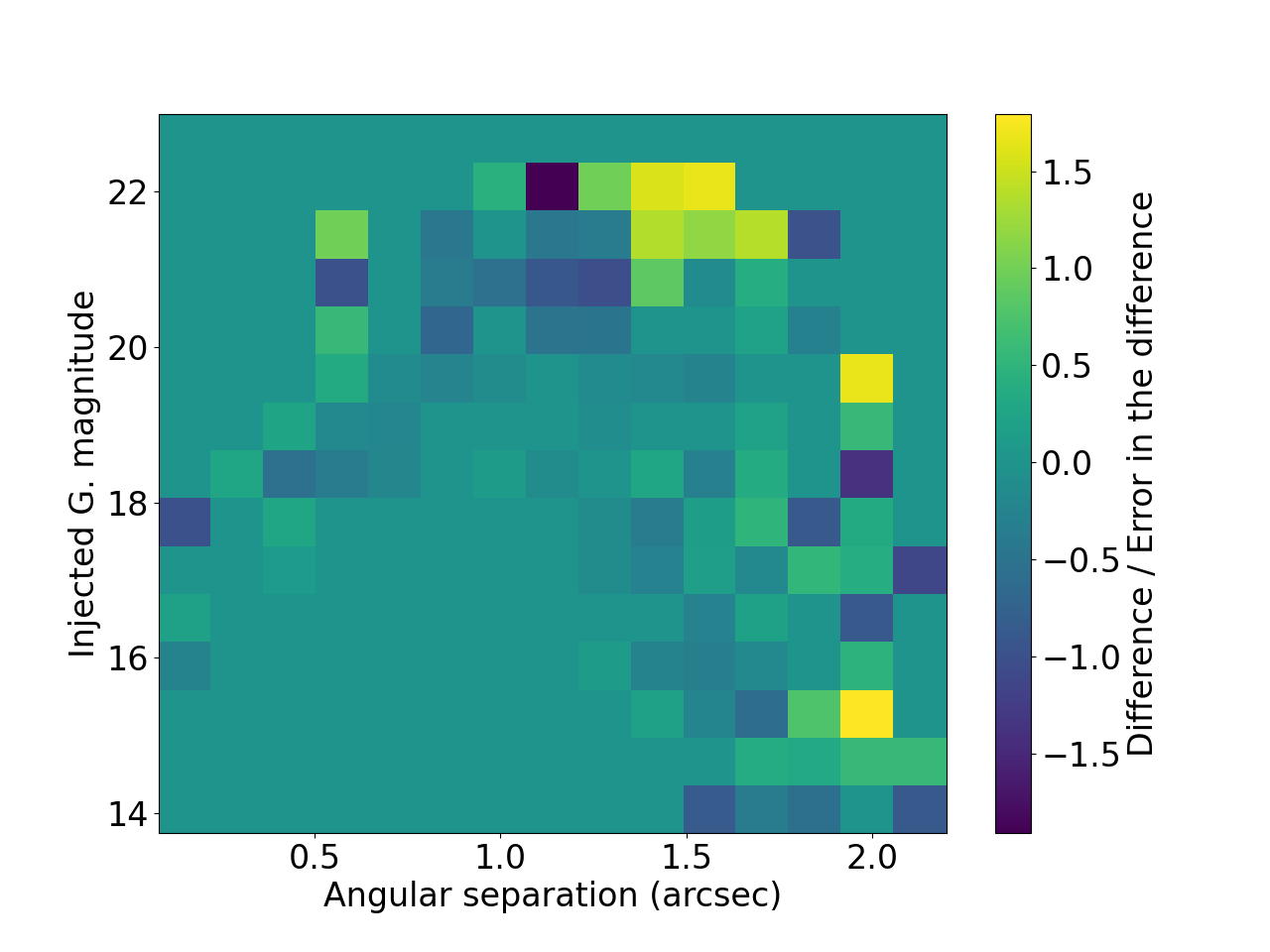}
\caption{Left: The fraction of the injected secondaries, injected into real data, which are recovered by the vanilla pipeline. Centre: The fraction of the injected secondaries, injected into mock data, which are recovered by the vanilla pipeline. Right: The significance of the difference (real-mock) between the recovery of secondaries injected into real and mock data.}
\label{fig:vanilla_results}
\end{figure*}
%
\begin{figure*}
\includegraphics[width=\textwidth/3]{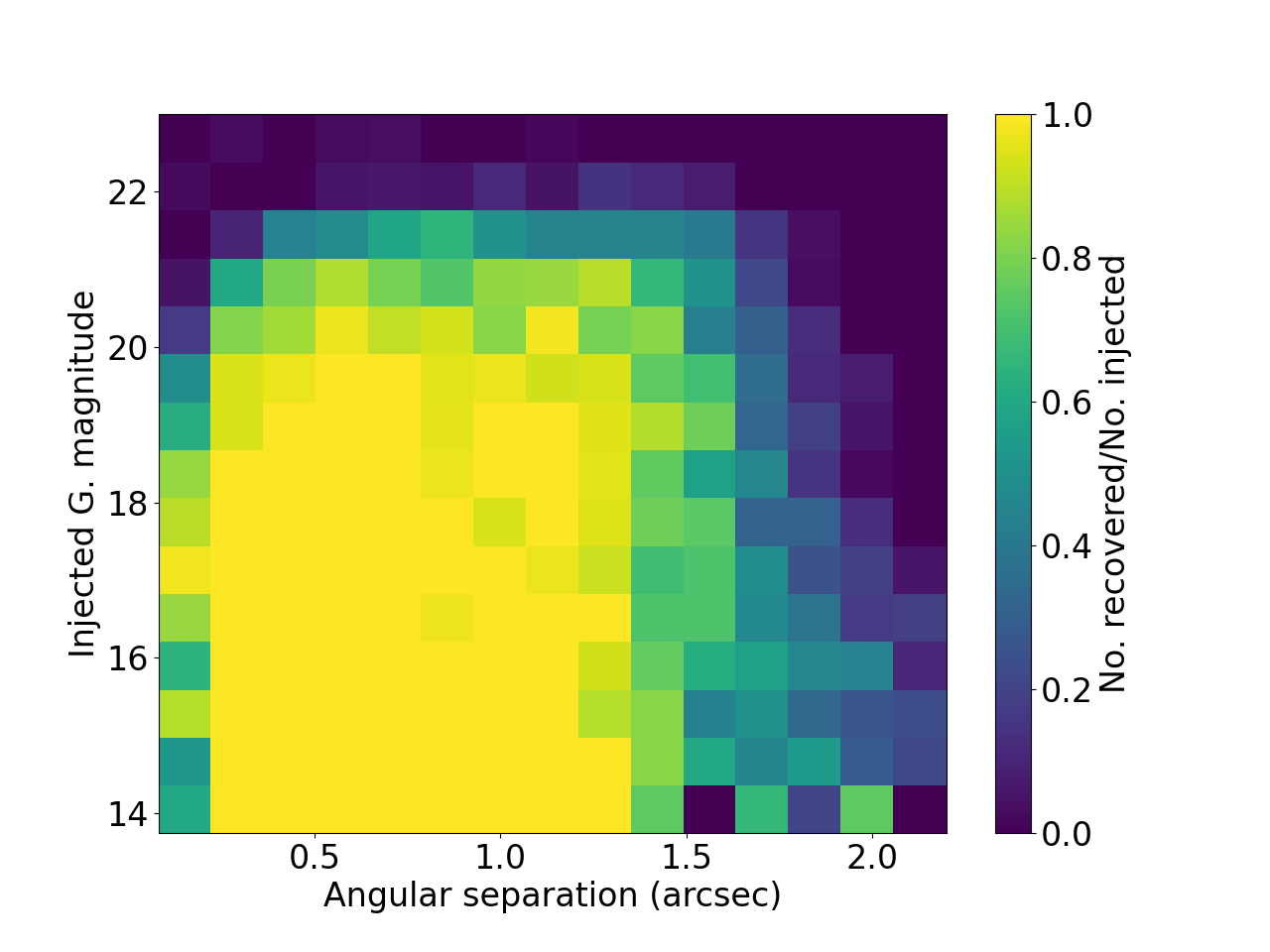}\includegraphics[width=\textwidth/3]{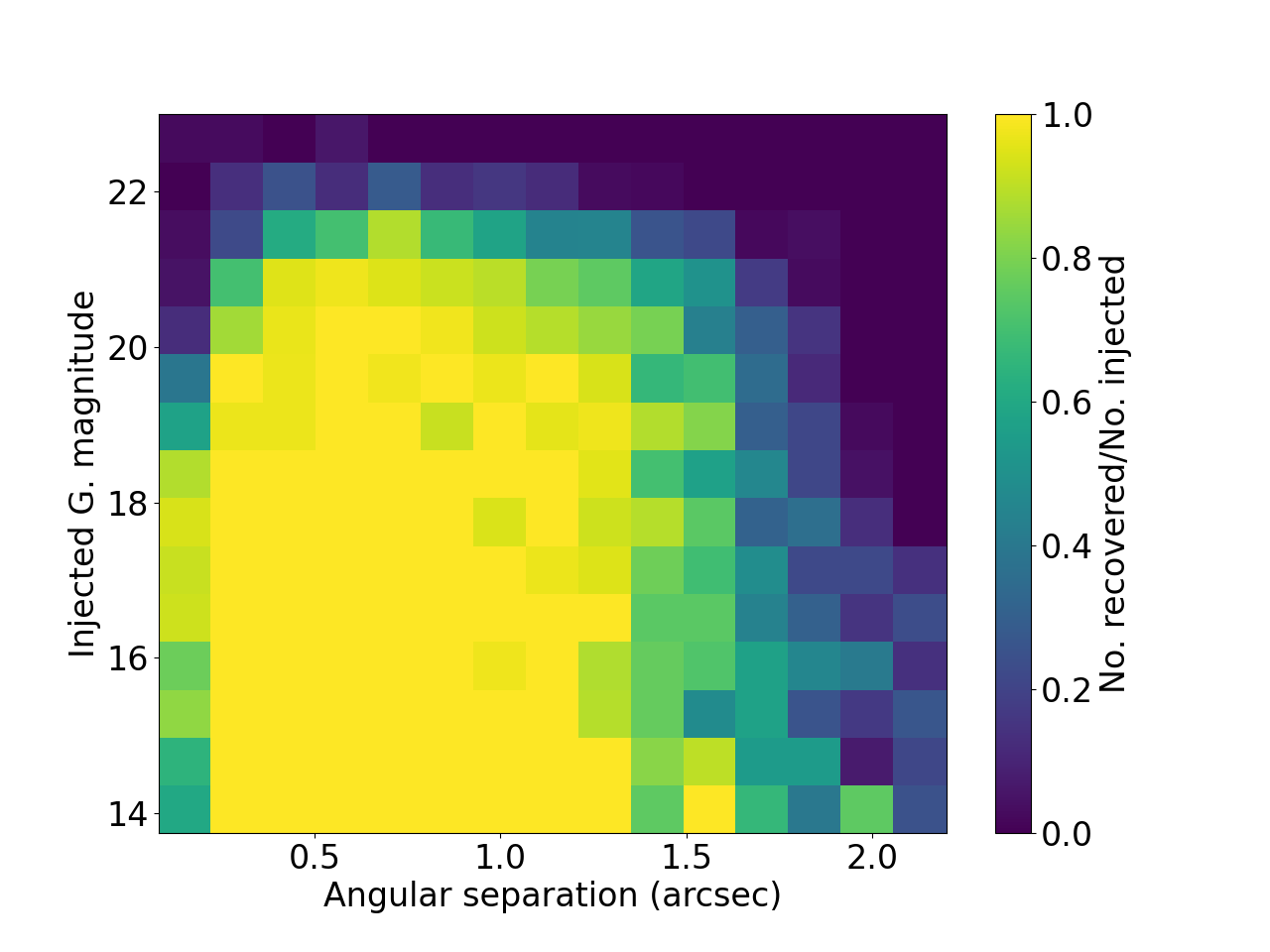}\includegraphics[width=\textwidth/3]{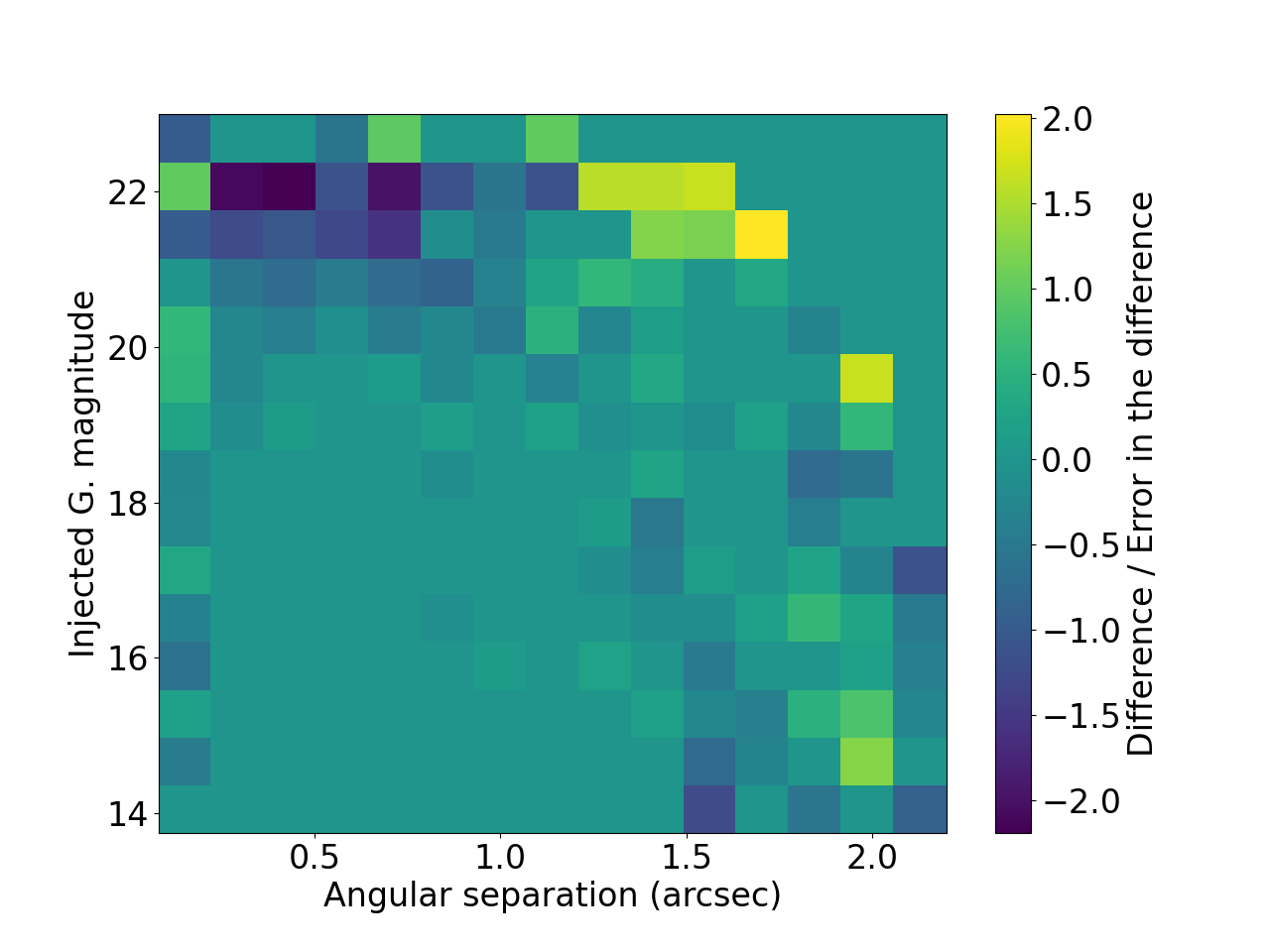}
\caption{Left: The fraction of the injected secondaries, injected into real data, which are recovered by the image-subtraction pipeline. Centre: The fraction of the injected secondaries, injected into mock data, which are recovered by the image-subtraction pipeline. Right: The significance of the difference (real-mock) between the recovery of secondaries injected into real and mock data.}
\label{fig:imgsub_results}
\end{figure*}
\begin{figure}
\includegraphics[width=\columnwidth]{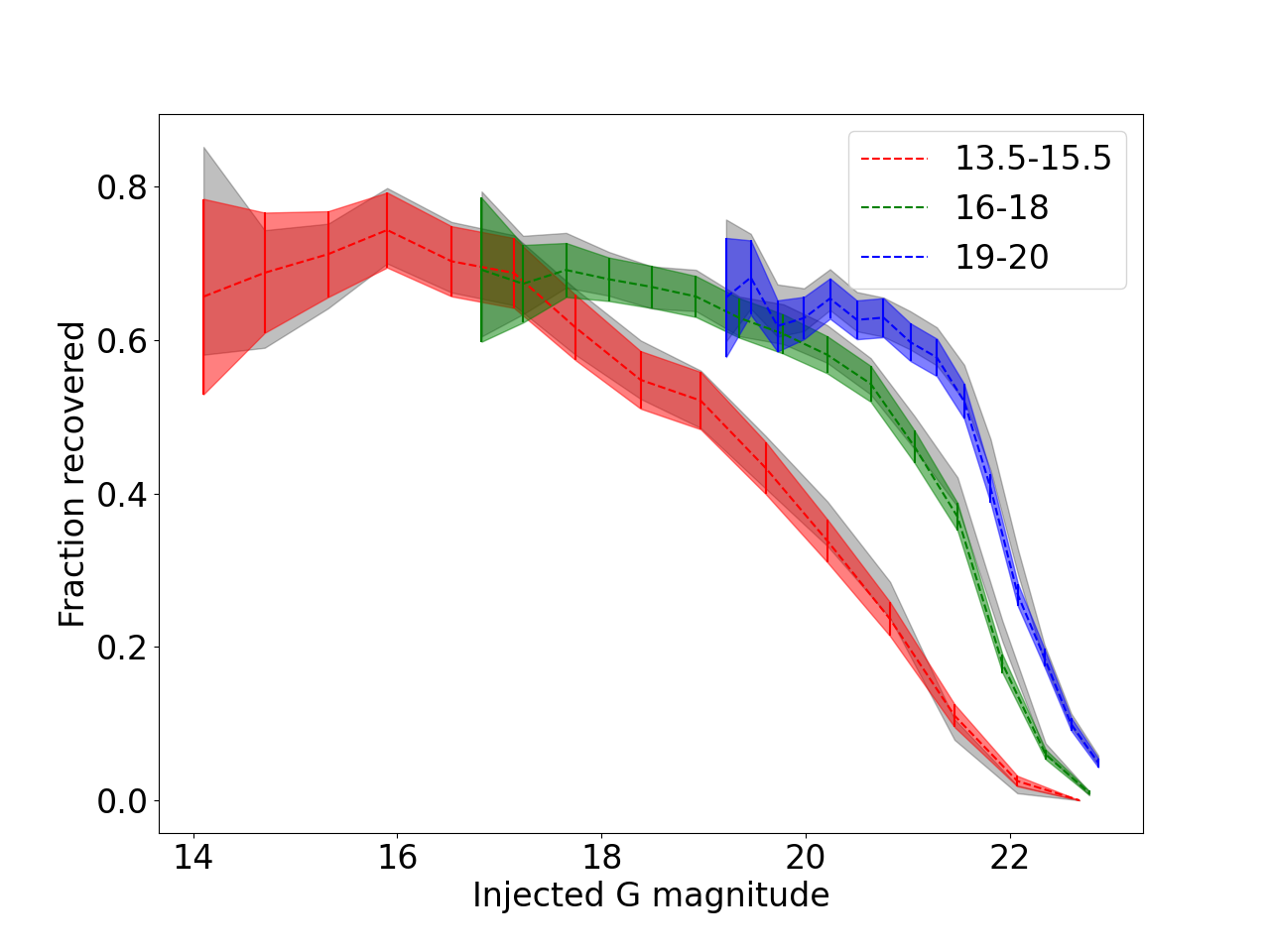}\\
\includegraphics[width=\columnwidth]{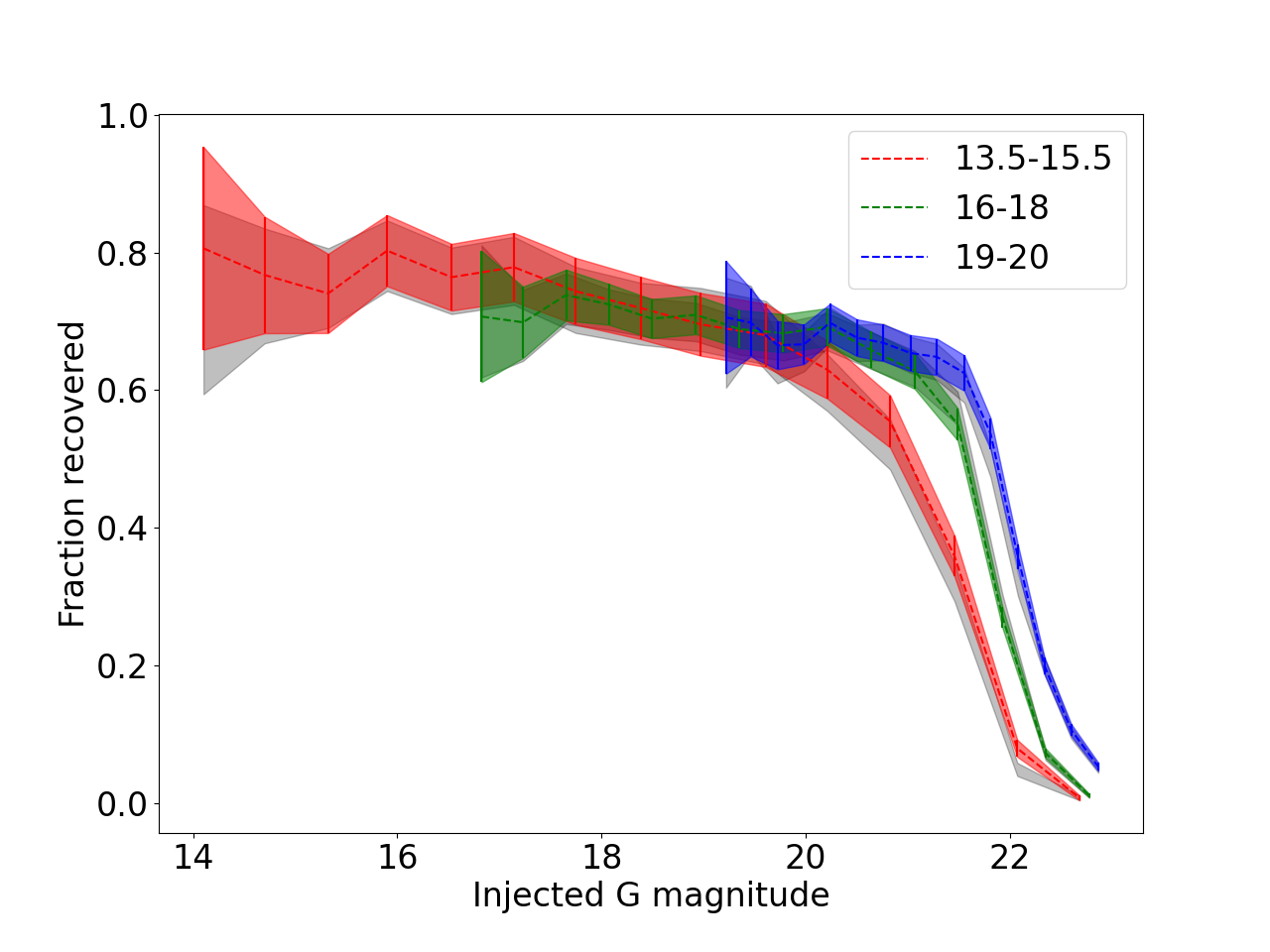}
\caption{The fraction of the sources which are recovered as a function of their injected $G$~magnitude.
Top:Vanilla pipeline. Bottom: Image-subtraction pipeline.
The colours correspond to the magnitude range of the primary sources into which the sources were injected. The shaded areas enclose the 1$\sigma$ errors in the recovered fraction. The shaded grey areas under the coloured areas correspond to injection into the real data, (the coloured area corresponds to injection into the mock data).
}
\label{fig:selection_function_results}
\end{figure}
\begin{figure}
\includegraphics[width=\columnwidth]{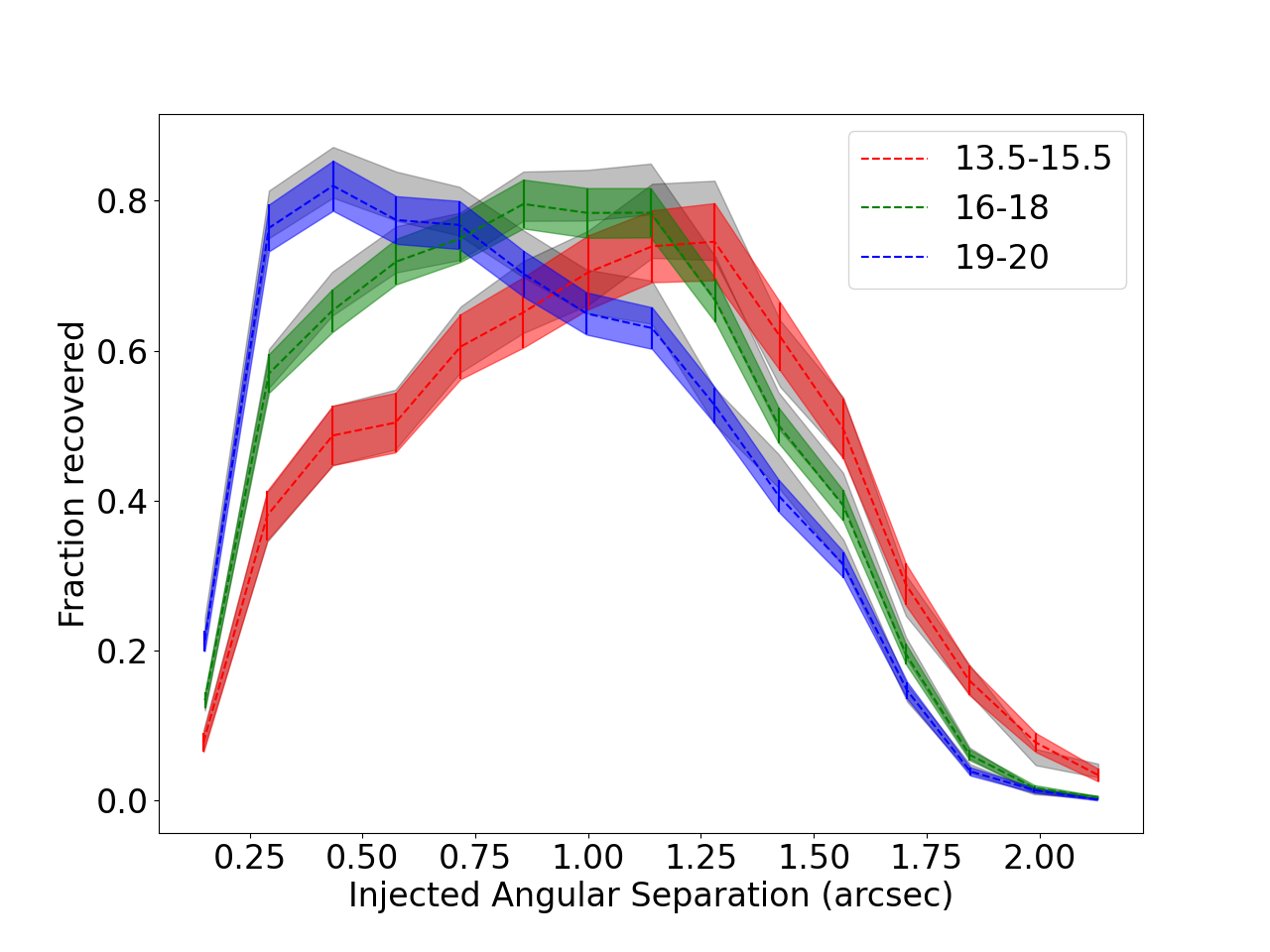}\\
\includegraphics[width=\columnwidth]{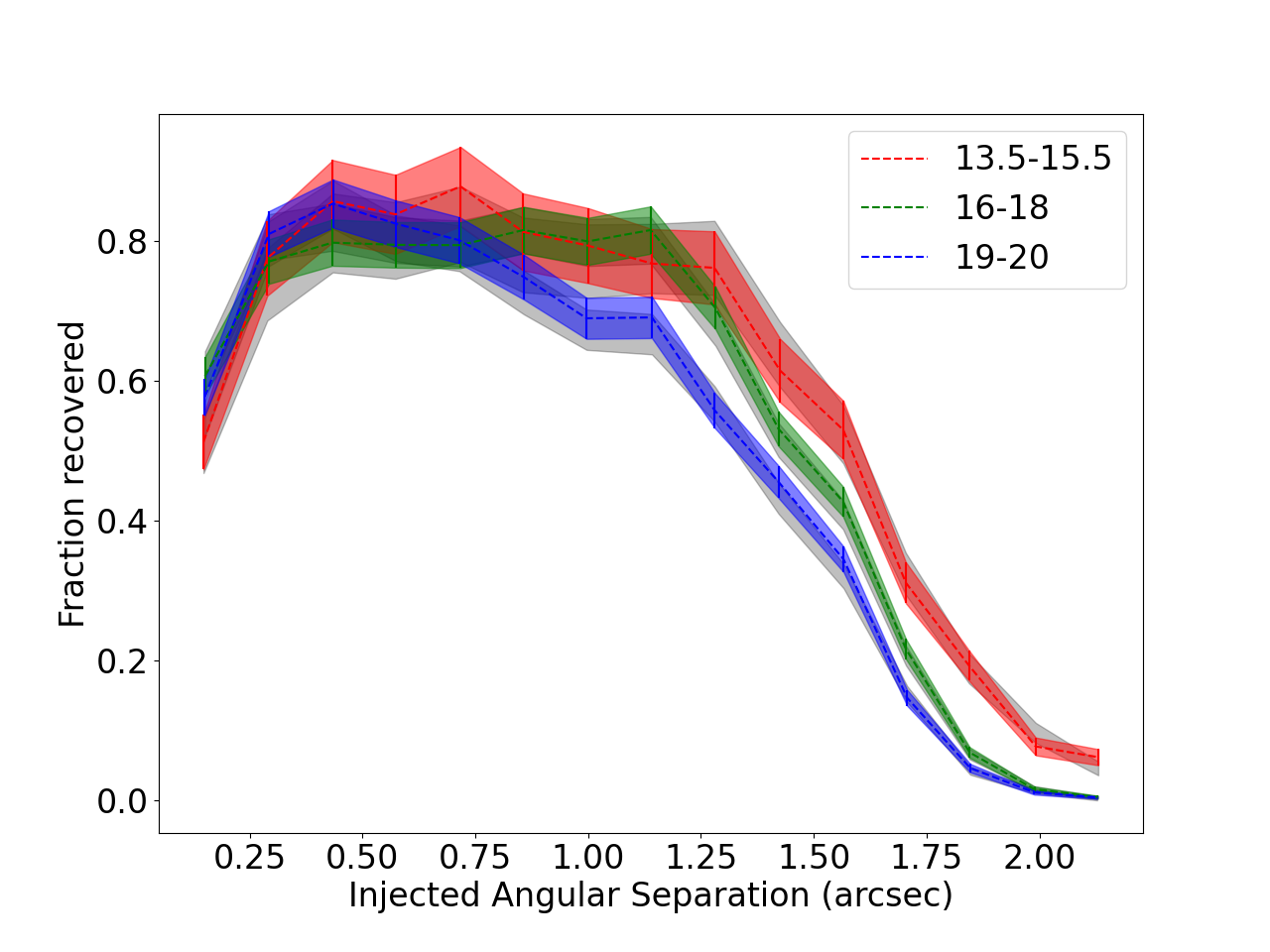}
\caption{The fraction of the sources which are recovered as a function of their injected angular separation from the primary source.
Top:Vanilla pipeline. Bottom: Image-subtraction pipeline.
The colours correspond to the magnitude range of the primary sources into which the sources were injected. The shaded areas enclose the 1$\sigma$ errors in the recovered fraction. The shaded grey areas under the coloured areas correspond to injection into the real data, (the coloured area corresponds to injection into the mock data).
}
\label{fig:selection_function_results_angsep}
\end{figure}
%
\begin{figure}
\includegraphics[width=\columnwidth]{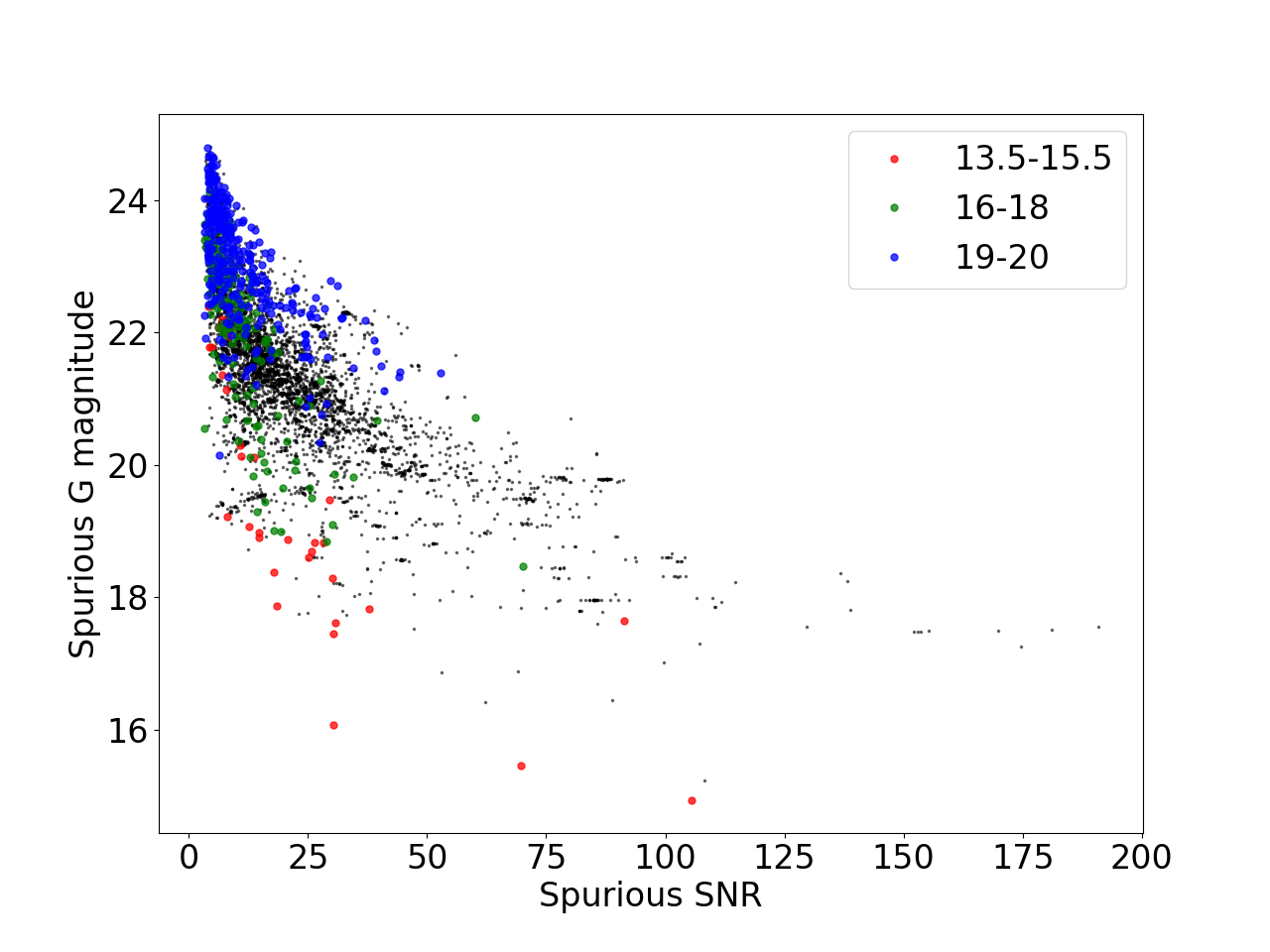}
\includegraphics[width=\columnwidth]{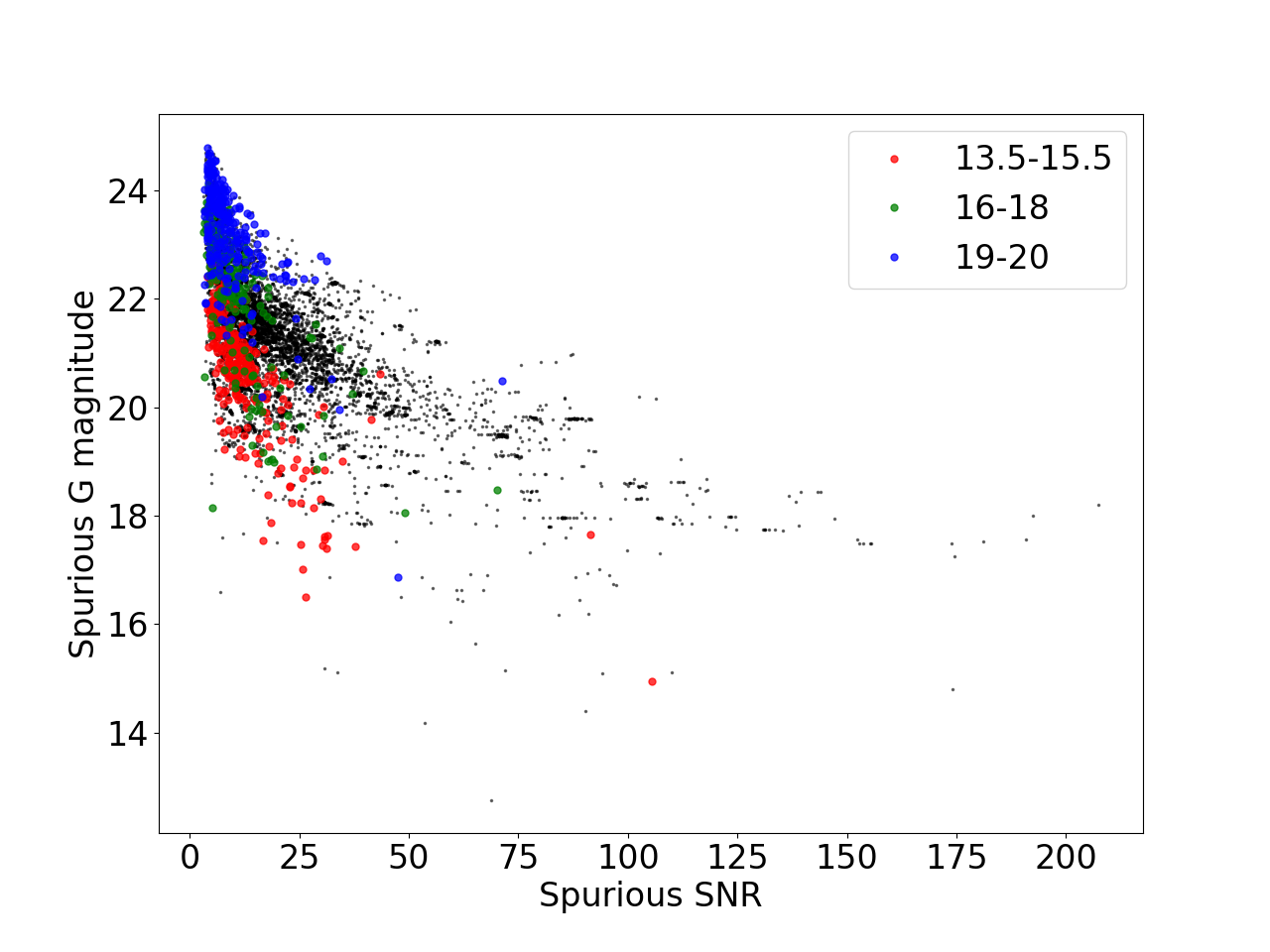}
\caption{The sources which were not matched either with the primary or the injected secondary source and hence deemed to be spurious. 
Top:Vanilla pipeline. Bottom: Image-subtraction pipeline.
The colours correspond to the magnitude range of the primary sources around which these spurious sources were found in the mock data. The black points are {\it spurious} sources found in the real data.
Many of these are likely to be real sources. The horizontal lines of dots are likely to be the same sources found on multiple occasions as the same primary source may be used more than once with a different injected secondary.
}
\label{fig:spurious_gmag_v_snr}
\end{figure}
\begin{figure}
\includegraphics[width=\columnwidth]{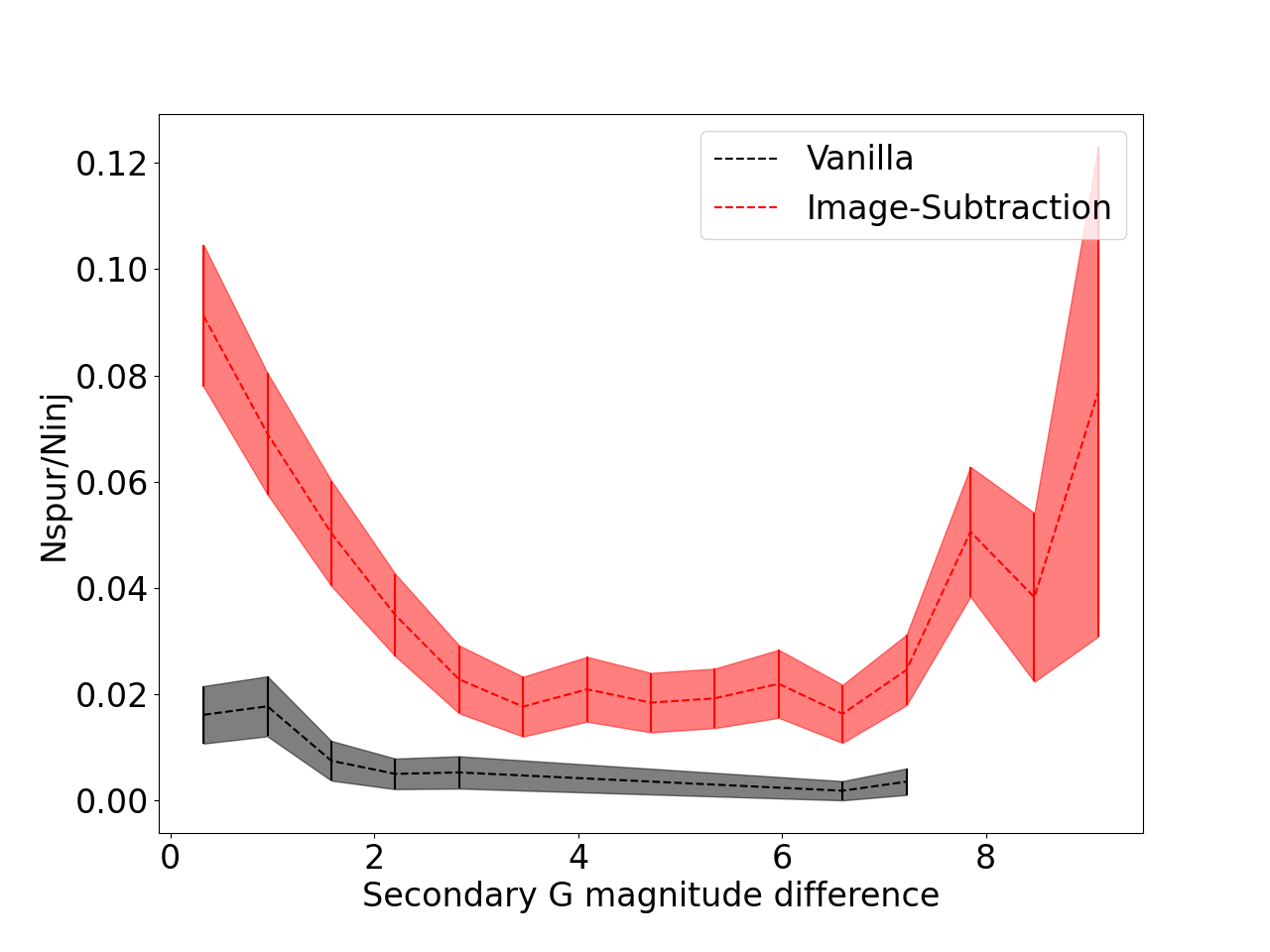}
\includegraphics[width=\columnwidth]{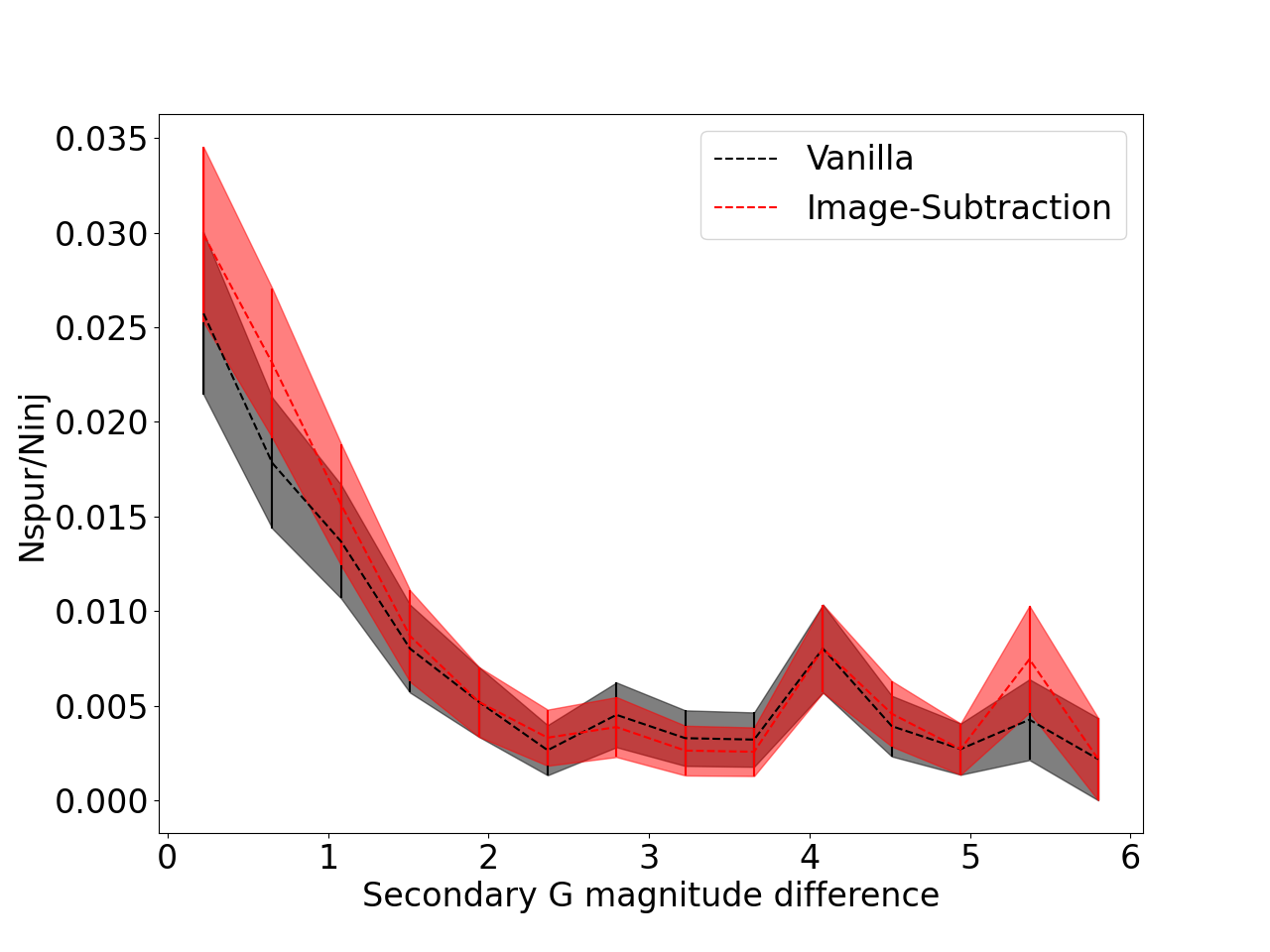}
\caption{The spurious rate, the number of spurious detections as a fraction of the number of secondaries injected, as a function of the magnitude difference between the primary and the injected secondary.
Top: primary sources in the $G$ magnitude range 13.5 to 15.5. 
Bottom: primary sources in the $G$ magnitude range 16.0 to 18.0.
The rates in both the image-subtraction and vanilla pipelines are shown. The shaded areas enclose the 1$\sigma$ errors in the spurious rate.
We see that there is an increase in the chance of a spurious detection, if there is a low magnitude difference companion.
}
\label{fig:spurious_v_magdif}
\end{figure}
\begin{figure}
\includegraphics[width=\columnwidth]{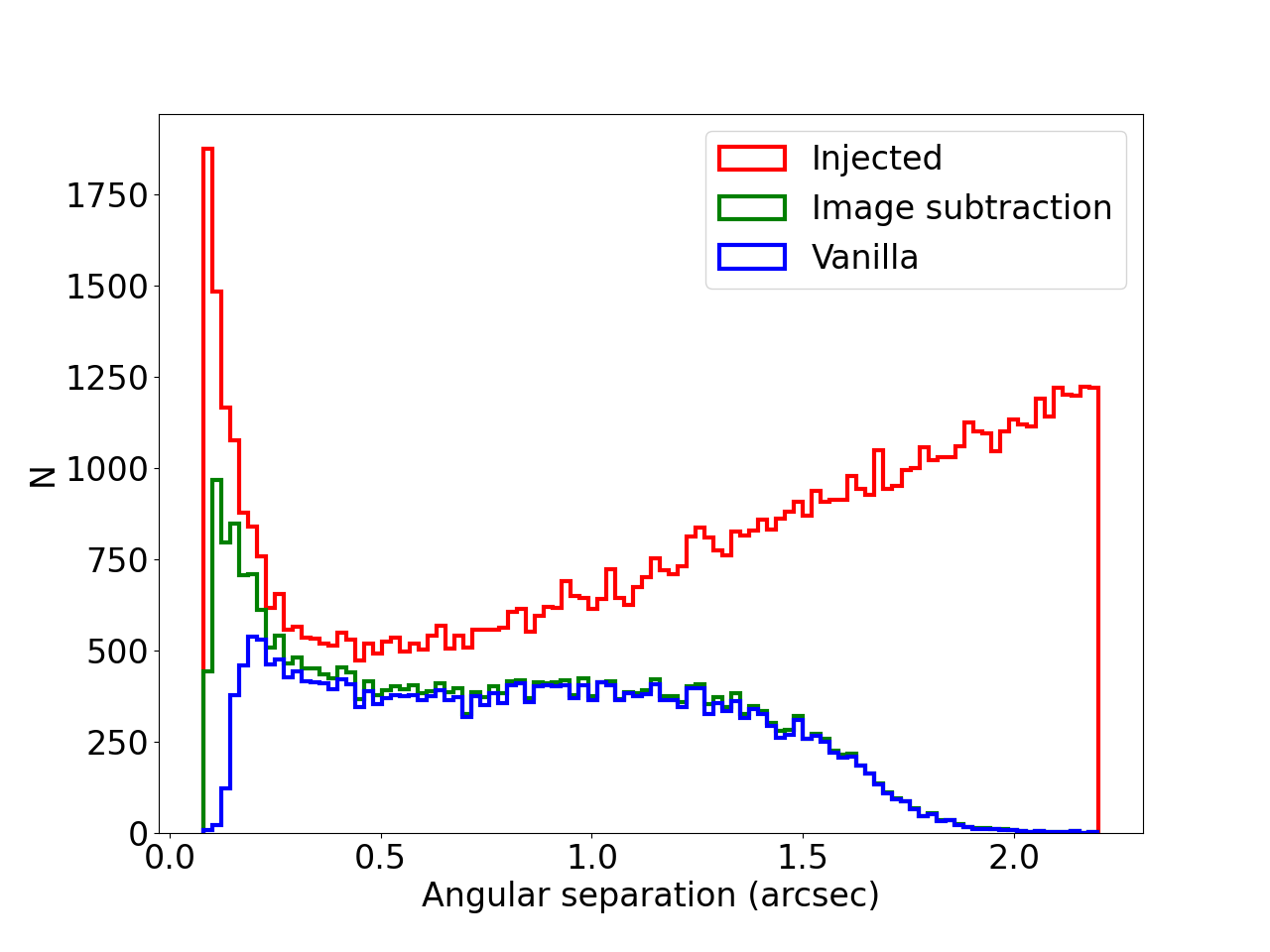}
\includegraphics[width=\columnwidth]{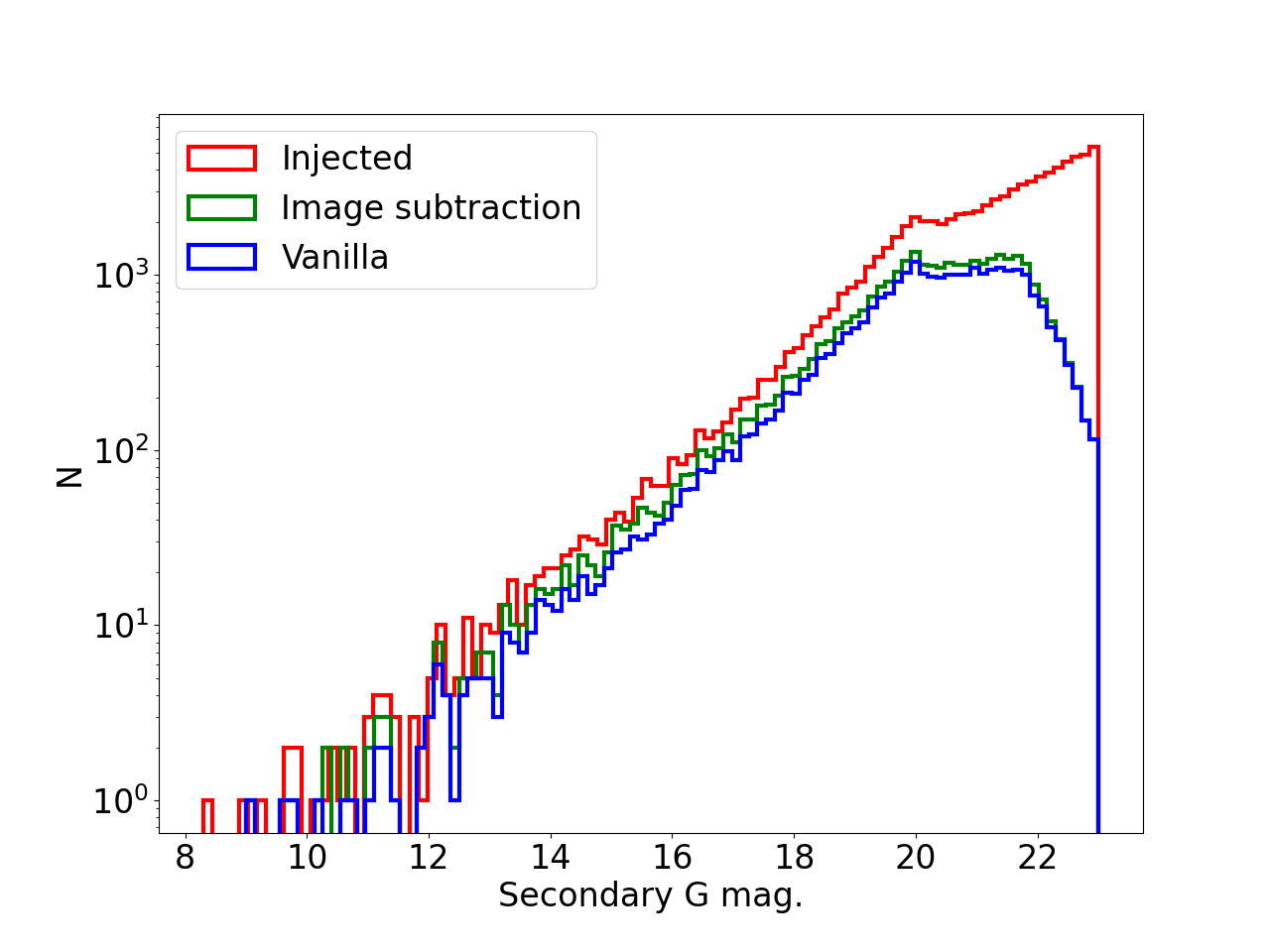}
\caption{The injected secondaries (drawn from the true underlying distribution of secondary sources) and those recovered by the vanilla and the image-subtraction pipelines as a function of their angular separation from the primary source (top) and their $G$ magnitude (bottom).
}
\label{fig:hists_true_dist}
\end{figure}

\begin{figure}
\includegraphics[width=\columnwidth]{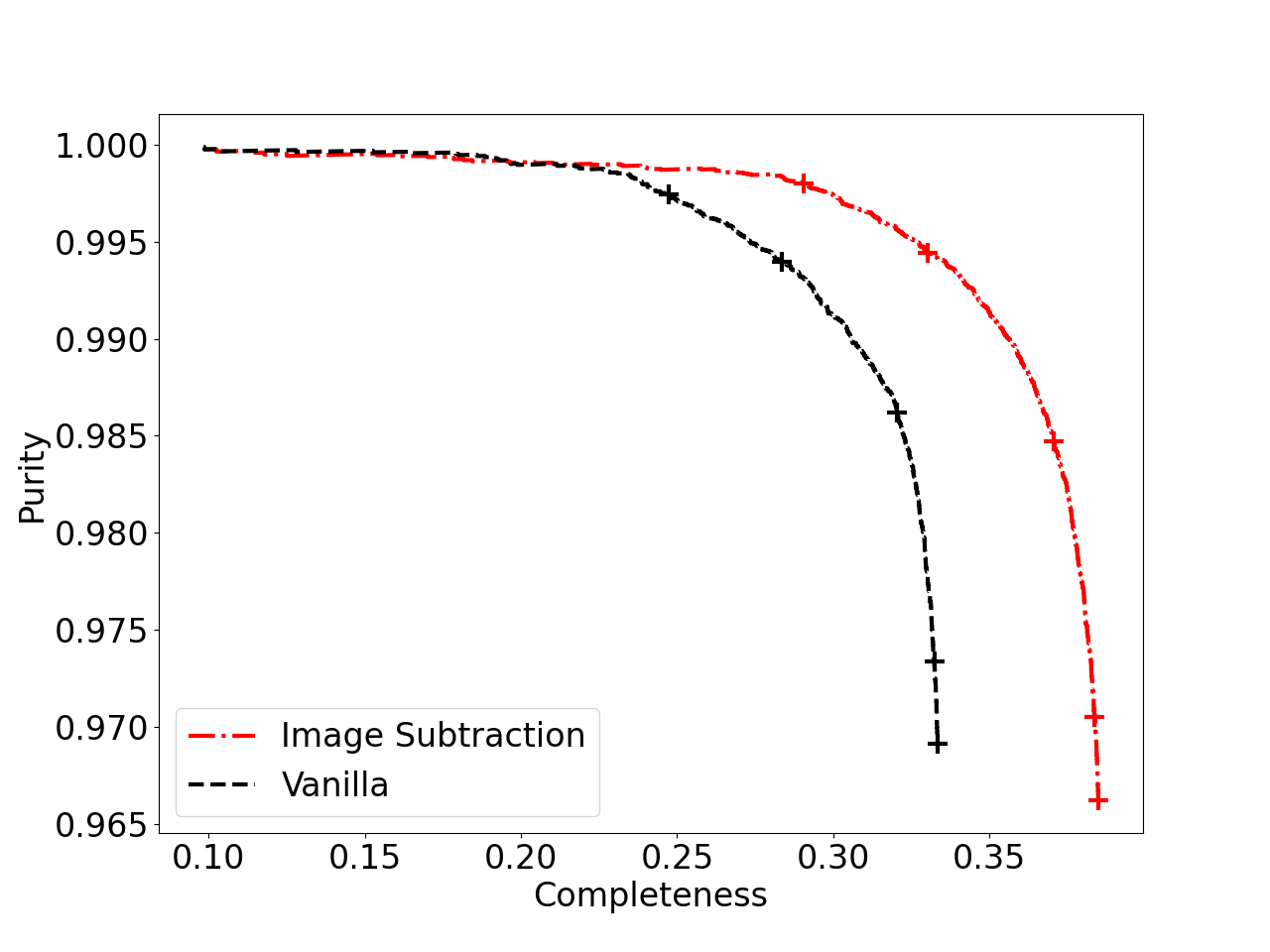}

\caption{The purity-completeness curves for the vanilla and image-subtraction pipelines.
These are evaluated by stepping through the data and decreasing the S/N threshold used for inclusion in the calculation of the ${\rm Purity} = N_{\rm real}/(N_{\rm real}+N_{\rm spurious})$ and ${\rm Completeness} = N_{\rm real}/N_{\rm injected}$.
The crosses from left to right on the curves correspond to S/N thresholds of 30, 20, 10, 5 and 3.
}
\label{fig:snr_curve}
\end{figure}
%
%
As was discovered in the attempt to create a list of known isolated sources, it was not possible to completely exclude the possibility of faint companions being present in the data.
The injection of companions into the real data will hence only allow the completeness and a worst case bound on the purity to be evaluated.
In order to make further progress in the characterisation of the purity some form of simulated data is required.
This simulated data should be as close to the real data as possible. The completeness derived from the real and simulated data can be used as a measure of their similarity. If the completenesses agree, the noise properties should be sufficiently similar that the level of spurious detections in the simulated data should reflect the actual level of spurious detections in the real data.
It should be noted, that this will provide a best case bound on the purity as there may be still be effects in the real data which result in spurious detections which have not been included in the simulations.

The simplest way to generate the required simulations is to replace the window samples of the real sources with samples simulated for a source of the same magnitude as the real source it replaces.
This removes the issue of faint companions in the data, and provides simulated data which is a direct match in terms of number of transits and orientation coverage and magnitude distribution as the real data.

The simulated data is generated using the data for a real source as follows:
\begin{itemize}
    \item{The position of the source at the epoch of the transit is determined from its astrometric parameters, adding   microarcsecond level random errors in the position at each epoch.}
    \item{The flux of the source (as found from its $G$-magnitude) is modified by the inverse of the calibration factor as provided by the Photometric One Day Calibration (PODC), for more information see Appendix~B of \cite{scienceAlerts}.}
    \item{Large residual background levels in the real data (caused by hot columns) are checked for and, if present, replicated in the simulated data}
    \item{A residual background is added, from a model which depends on the AF CCD}
    \item{Poisson Noise is added}   
\end{itemize}
The simulated (mock) data may then be treated in the same way as the real data.

The real and the mock data were compared over three different magnitude ranges.
The primary sources into which the secondary sources are injected are chosen such that they have the same distribution in ecliptic latitude and in $G$-magnitude (for the magnitude range in question) as the \gaia{} DR3 catalogue \cite{GaiaDR3}.
The ecliptic latitude constraint is necessary as the coverage of the transits over the source varies as a function of ecliptic latitude and this coverage can affect the chance of a spurious detection (sources with poor coverage in the orientation of transits are more susceptible).

Figure~\ref{fig:numberInjectedSources} shows the number of injected sources as a function of their $G$-magnitude and their angular separation from their primary source. The injected sources are drawn from a uniform distribution in angular separation (0.08 to 2.2 arcseconds) and magnitude difference (0.01 to 12.0) from their primary source.
The injected source is always fainter than its primary, though it is never fainter than $G=23.0$ (another draw in the magnitude difference is performed in these cases).
The population of primary sources into which these sources are injected are all in the $G$-magnitude range 13.5 to 15.5. The distribution of the number of primary sources at each magnitude in this range follows that of the \gaia{} catalogue ($N \propto G_{\rm mag}^{\alpha}$, where $\alpha=10.6$).
This is the reason for the small number of injected sources brighter than $G\sim15$ as they must be fainter than their primary source, and there are fewer bright primary sources than faint ones in this data-set.

This simulated data and the corresponding real data used to generate it, is injected with the same secondary sources and then processed using both the vanilla and image-subtraction pipelines.
The results of this analysis are shown in Fig.~\ref{fig:vanilla_results} for the vanilla pipeline and Fig.~\ref{fig:imgsub_results} for the image-subtraction pipeline, for the $G$-magnitude range 13.5 to 15.5.
In both figures the fraction of injected companions which are detected in both the real and the mock data as a function of their $G$-magnitude and angular separation from their primary are shown. 
Additionally, in each figure the significance of the difference between the recovery in real and mock data is shown.
These figures show that the recovery of companion sources is similar between the real and the mock data, for both the vanilla and image-subtraction pipelines.
They also show the gains in the image-subtraction pipeline over that of the vanilla in the detection of fainter companions closer to their primaries.
The drop in the fraction of sources recovered at larger angular separations in Figs.~\ref{fig:vanilla_results}~and~\ref{fig:imgsub_results} can be understood by the window sizes and the choice in the size of the reconstructed images.
Windows are rectangular and for AF windows wider in AC than in AL, with the reverse being true for SM windows. The AC size for all windows (SM and AF) is always $\sim2$ arcseconds regardless of the window class. 
Hence, for angular separations beyond $\sim1$ arcsecond there is no contribution from the AF data (the windows are centred on the primary source), and there is progressively less overlap in the SM windows as the angular separation increases.
Hence, at larger angular separations it is much more likely that any source detectable in the image has already been discovered by the on-board detection algorithm.
Given the image reconstruction algorithm scales non-linearly with the number of pixels in the output image, it makes sense to restrict the size of the reconstructed image due to the diminishing returns and increasing computational expense.
All reconstructed images are square with a side length of 3 arcseconds. Once secondary sources are further than $\sim1.5$ arcseconds from the primary source, they may not appear in the reconstructed image depending on their orientation with respect to it.
It is this effect which determines the abrupt drop off fraction of secondaries recovered at angular separations beyond $\sim1.5$ arcseconds.

Figures~\ref{fig:selection_function_results}~and~\ref{fig:selection_function_results_angsep} show the results for all three magnitude ranges investigated. Here the fraction of the injected sources which are found is shown as a function of their injected $G$ magnitude and angular separation from their primary, respectively. The colours correspond to the magnitude range of the primary sources into which the sources were injected, and the shaded areas corresponds to the $1\sigma$ errors on the recovered fraction found from the mock data. The shaded grey areas correspond to injection into the real data.
These figures show good agreement between the recovery of sources injected into the real and the mock data for all three magnitude ranges.
The improvement in the recovered fraction when using the image-subtraction pipeline is seen to depend on the magnitude range of the primary sources, with the biggest gains for the brightest magnitude range.
This can be understood as the image-subtraction pipeline improving the contrast of the fainter, closer sources and allowing their detection, but there is still a limit to how faint a source may be detected.
For the fainter magnitude ranges of primary sources, proportionally more of the secondary sources bright enough to be detectable are already detectable by the vanilla pipeline (having sufficiently small enough magnitude differences from their primary).
This then reduces the gains seen in the image-subtraction pipeline. 

Agreement in the recovery of injected sources from the real and mock data using the vanilla and image-subtraction pipeline means that the spurious detections in the mock data should be representative of the spurious population in the real data.
While this is not strictly true, this represents the best available estimate of the spurious population and hence the reliability of the SEAPipe pipelines.
Figure~\ref{fig:spurious_gmag_v_snr} shows the spurious detections found in the vanilla and image-subtraction pipelines for both the real and mock data-sets.
The spurious detections were classified as such as they were not matched to the primary or the injected companion source.
However, in the real data there are known to be residual real sources, as described in Sect.~\ref{sec:isolated}.
In Fig.~\ref{fig:spurious_gmag_v_snr} the nominally spurious detections from the real data (recall that some of these {\it spurious} detections are real faint companion sources) can be seen to occur at higher S/Ns than the mock data, where the S/N for a source is the ratio of the flux and its error returned by the image parameter analysis step.
Also noticeable are horizontal lines of dots of a group of {\it spurious} detections with the same $G$ magnitude over a narrow range of S/Ns. 
These are likely to be the same source found on multiple occasions around the same primary as a given primary source may be used more than once with a different injected secondary.
The overall form of spurious is similar between the real and the mock data-sets, however, as expected.

The increase in the number of spurious detections found in the image-subtraction pipeline for the brightest primary magnitude range is noticeable in Fig.~\ref{fig:spurious_gmag_v_snr}, while no increase is immediately obvious for the other two primary magnitude ranges.

The rate of spurious detections has also been investigated as a function of the $G$ magnitude of the injected companion.
The long-rectangular samples containing the peak flux from the primary and the companion source in transits crossing the primary in different orientations may overlap. This can produce a region of excess flux in the images (especially in cases of poor coverage) which may be detected as an additional source.
The chance of a spurious detection therefore may increase if there is a companion source present, particularly a brighter one.
Figure~\ref{fig:spurious_v_magdif} shows the results of an investigation into the chance of a spurious detection as a function of the difference between the $G$ magnitude of the injected companion and its primary.
The number of spurious detections found in mock data with an injected companion of a given $G$ magnitude difference from its primary is divided by the total number of injected companions with that $G$ magnitude difference.
This provides the chance of a spurious detection being found as a function of the magnitude difference between the primary and the injected companion.
In Fig.~\ref{fig:spurious_v_magdif} we can see that this is indeed larger for smaller magnitude differences; and that this is true for both the vanilla and image-subtraction pipeline.
This figure also shows the large increase in the chance of a spurious detection in the image-subtraction pipeline in the 13.5-15.5 primary magnitude range.
The majority of these are due to artefacts induced by the presence of a secondary source, but the rate beyond a magnitude difference of $\sim 4$ (to $\sim 7$) is consistent with the rate of spurious detections around isolated primary sources in this magnitude range with no injected secondaries.
However, we saw in Figs~\ref{fig:selection_function_results}~and~\ref{fig:selection_function_results_angsep} that the image-subtraction pipeline retrieves a substantially larger fraction of the injected sources than the vanilla pipeline in this primary magnitude range.
An assessment of the completeness versus the purity is needed in order to find the best performing pipeline.

In our comparison of the performance of the pipelines on real and mock data we have computed the selection function (Figs~\ref{fig:vanilla_results}~and~\ref{fig:imgsub_results}). 
This selection function expresses the probability of detecting a secondary source with a particular angular separation and magnitude difference from the primary source in a given magnitude range. 
The selection function does not depend on the distribution of the secondary sources in the angular separation and magnitude difference parameter space.
It served to demonstrate that the mock data is sufficiently close to the real data to be used to assess the performance of the vanilla and image-subtraction pipelines in the generation of the catalogue of secondary sources produced by SEAPipe. 
The performance of these pipelines, however, should to be judged on the completeness and the purity of the resultant catalogue.
The completeness may be evaluated as:
\begin{equation}
    {\rm Completeness} = \frac{N_{\rm real}}{N_{\rm injected}}
    \label{eqn:completeness}
\end{equation}
and the purity as:
\begin{equation}
    {\rm Purity} = \frac{N_{\rm real}}{N_{\rm real}+N_{\rm spurious}}
     \label{eqn:purity}
\end{equation}
where $N_{\rm real}$ and $N_{\rm spurious}$ are the number of real and spurious detections made by the pipeline, and $N_{\rm injected}$ is the number of injected sources.
In order for this completeness to reflect the completeness of the catalogue produced from the analysis of the real \gaia{} data, the properties of the injected sources should match those of the secondary sources in the \gaia{} data.
Hence, the injected sources must be drawn from the true underlying distribution of secondary sources as a function of their angular separation and magnitude difference from their primary source, including how this varies with the magnitude of the primary source.
While the selection function may be used to compute the completeness given this true underlying distribution, we also wish to know about the purity. 
As seen previously, in Fig.~\ref{fig:spurious_v_magdif}, the probability of a spurious detection increases given the presence of a low magnitude-difference secondary.  
It is clear therefore, that the purity also depends on the true underlying distribution of secondary sources.
Hence, in order to assess both the completeness and purity it is necessary to perform another Monte Carlo simulation, injecting secondary sources drawn from this distribution.
We used the universe model of \cite{robin2012} to compute the underlying distribution of secondary sources.
These secondary sources are a combination of chance projections and bound companions, where the fraction of bound companions to chance projections varies as a function of the magnitude of the primary source, with the fraction of bound companions decreasing for fainter primaries.

Figure~\ref{fig:hists_true_dist} shows histograms in magnitude and angular separation of the injected sources (recall the secondaries are by definition fainter than their primaries), in these simulations using mock data.
This set of simulations only restricted the primary magnitude range to those brighter than $G\approx20$, and selected the primaries such that they follow the magnitude distribution seen in the \gaia{} catalogue.
The secondaries are restricted to those brighter than $G\approx23$.
In Fig.~\ref{fig:hists_true_dist}, the peak of sources at small angular separations is due to the bound companions, with the increasing slope towards larger angular separations being due to chance projections.
Again we see the impact of the image-subtraction pipeline at smaller angular separations, with the additional sources discovered by this pipeline primarily being from the bound component.
In the bottom panel of Fig.~\ref{fig:hists_true_dist}, the change in the slope of the injected secondaries at $G \approx 20$ is a result of the $G \approx 20$ magnitude of the faintest primary source into which secondaries are injected, and as in the previous Monte Carlos runs no secondaries fainter than $G=23$ are injected.

The curves in Fig.~\ref{fig:snr_curve} show the completeness and purity of the resultant catalogues of secondary sources made by the vanilla and image-subtraction pipelines. 
These curves are computed by ordering all the detections as a function of decreasing S/N and evaluating eqs~\ref{eqn:completeness}~and~\ref{eqn:purity} as each additional detection is added to the catalogue.
Hence, the S/N threshold required for the inclusion of a source in the catalogue decreases along the curves from left to right. 
The crosses in Fig.~\ref{fig:snr_curve} indicate the purity and completeness of the catalogue for given S/N thresholds, in this case 30, 20, 10, 5 and 3.
At high S/N thresholds both pipelines are pure but very incomplete.
As the S/N threshold for inclusion in the catalogue is reduced, the completeness of both the vanilla and image-subtraction catalogues increases while the purity decreases.
However, for a given purity the image-subtraction catalogue is now always more complete than the vanilla catalogue.
For example a catalogue with a purity of 0.99 (99\,\% real, 1\,\% spurious) made with the image-subtraction pipeline is 35.6\,\% complete down to $G=23.0$, while the vanilla pipeline is only 30.6\,\% complete.
The S/N thresholds required to produce these 99\,\% pure catalogues are 14.0 for the image-subtraction pipeline and 14.2 for the vanilla pipeline.
This shows that while the image-subtraction pipeline may result in more spurious detections (Fig.~\ref{fig:spurious_v_magdif}), the number of additional real detections is significantly greater.
The importance of using the expected distribution for secondary sources in this analysis is revealed as the bound component at small angular separations from the primary source forms the majority of these additional sources found by the image-subtraction pipeline (Fig.~\ref{fig:hists_true_dist}).
This result should be representative of the result of the analysis of the real \gaia{} data, given this simulation used the expected magnitude distribution of primaries and magnitude and angular separation distributions for the injected secondaries.
%
\begin{figure}
\includegraphics[width=\columnwidth]{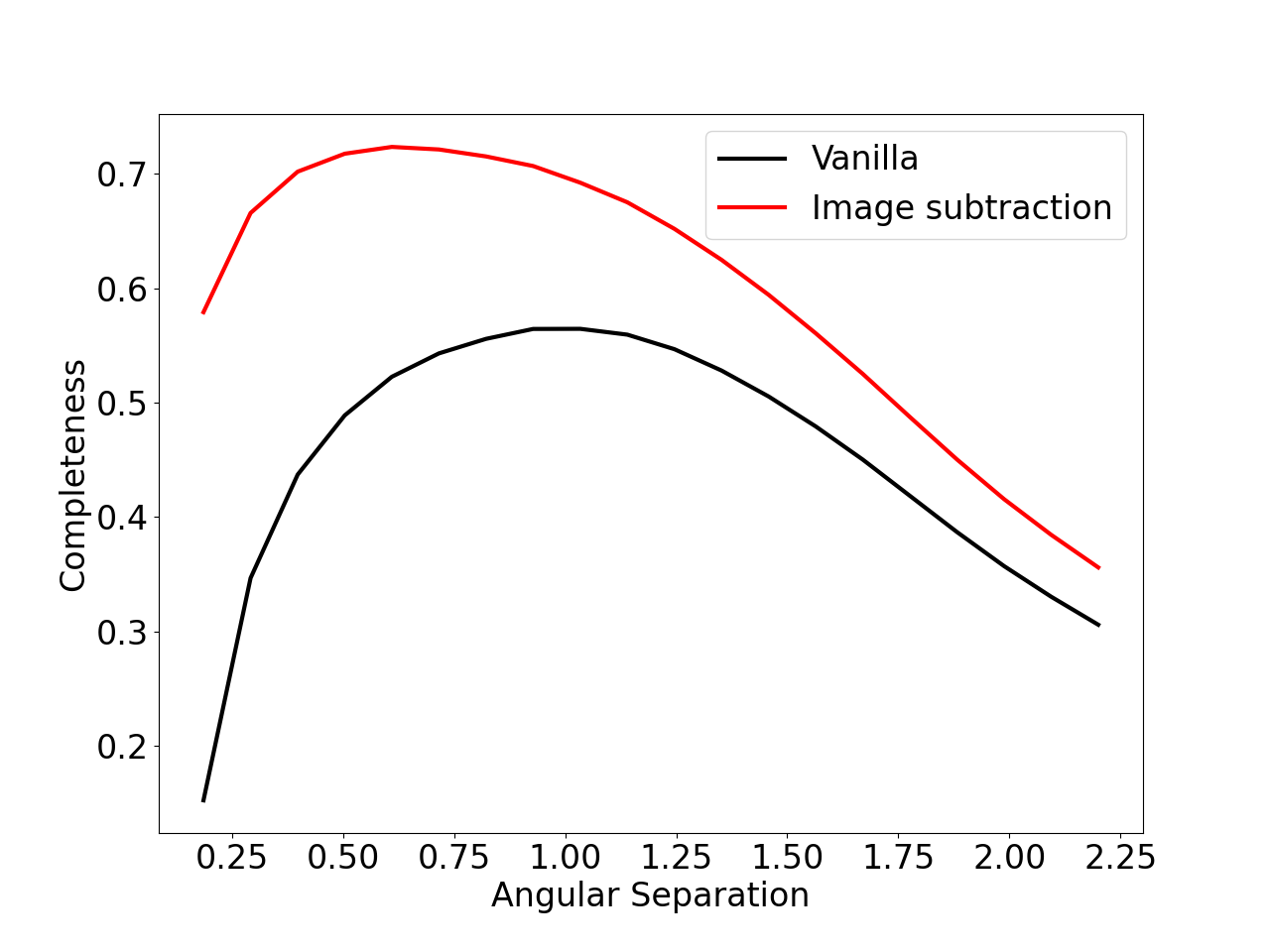}

\caption{The completeness of the vanilla and image-subtraction pipelines, for secondary sources within a given angular separation of the primary source.
These are for the 99\,\% purity catalogues obtained by cutting the catalogues at an S/N of 14.0 for image subtraction pipeline and an S/N of 14.2 for the vanilla pipeline. 
}
\label{fig:within_angsep}
\end{figure}
The low level of completeness may be understood from the distribution of injected sources and the size of the reconstructed images.
Recall that the reconstructed images are square with sides of 3 arcseconds in length, so that secondary sources more than $\sim 1.5$ arcseconds away may not appear in the image. The injected secondaries with the largest angular separations ($\sim 2.2$ arcseconds) may only appear in the image at a corner; in any other location they will be outside it.
Additionally, the model from which the properties of the secondary sources are drawn is a combination of chance projections and bound companions. The number of chance projections increases with angular separation, with the increasing area available in which a source could occur.
This means there are many injected sources with larger angular separations, as seen in the top panel in Fig.~\ref{fig:hists_true_dist}, many of which will not be detectable, this will obviously impact the resultant completeness.
The chance projections are also more likely to have larger magnitude differences from their primary source due to the larger number of fainter sources, making them less detectable. This will also impact the completeness.
To illustrate these effects, Fig.~\ref{fig:within_angsep} shows the completeness for secondary sources within a given angular separation of the primary source, for the 99\,\%  purity catalogue described above.
Here we see that the image subtraction pipeline finds 72.3\,\% of all secondaries within 0.6 arcseconds of the primary sources, while the vanilla pipeline retrieves 56.4\,\% of all secondaries within 1.0 arcsecond, both down to G=23.0.
Beyond these angular separations the impact of the detectability of the chance projections starts to be noticeable and the completeness decreases.

The stated aim of SEAPipe is to find those sources not found by the on-board detection algorithm which could otherwise perturb the results if they are not properly accounted for.
The reduced sensitivity at large angular separations is not unexpected or concerning in this respect as these sources are likely to have been seen by the on-board detection, or be faint enough to be below its detection threshold.
The main contribution of SEAPipe will be the detection of new sources at angular separations less than $\sim 0.5$ arcseconds, and their subsequent addition to the \gaia{} catalogue.

\section{Discussion}
\label{sec:discussion}

This paper has presented the machinery required to produce a catalogue of secondary sources found in the \gaia{} data with a given level of purity.
Exactly what level of purity should be targeted for this catalogue, however, is beyond the scope of this work.
These sources are found in the images reconstructed by combining the transit data of individual \gaia{} sources, and many will have been missed by the on-board detection and hence would otherwise not be present in the final \gaia{} catalogues.
There are two potential pipelines which may be used, the image-subtraction and vanilla pipelines. The image-subtraction pipeline is superior in terms of completeness for a given level of purity. However, this comes at a larger computational cost. 
Given that the majority of the secondary sources found only in the image-subtraction pipeline are bound companions, the chance of an additional source being found (i.e. only in the image-subtraction pipeline) for any individual \gaia{} source decreases with increasing magnitude.
However, as the number of sources at each magnitude increases as the sources become fainter the resultant number of additional sources in the catalogue found by the image-subtraction pipeline increases with increasing primary source magnitude. 
In other words, there are in total more additional sources found around fainter stars, simply because there are more fainter stars in the \gaia{} catalogue.
Hence, ideally the image-subtraction pipeline should be run on all \gaia{} sources. If, however, there are insufficient computational resources available then the brighter sources should be prioritised for processing with this pipeline.
Once a pipeline or combination of pipelines (based on the available resources) is chosen, SEAPipe can be tested using the analysis described in Sect.~\ref{sec:mcqa} to evaluate the completeness and purity for the resultant catalogue. The desired level of purity will then be used to set the S/N of the sources which will be accepted into the catalogue of secondary sources.
%

\begin{acknowledgements}
This work has made use of data from the ESA space mission Gaia, processed by the Gaia Data Processing and Analysis Consortium (DPAC). Funding for the DPAC has been provided by national institutions, in particular the institutions participating in the Gaia Multilateral Agreement.
This work has been financially supported by the United Kingdom Space Agency, through the following grants to the University of Cambridge, 
PP/D006546/1, ST/1000542/1, ST/K000756/1, ST/N000641/1, ST/S000089/1, ST/W002469/1, ST/X00158X/1.
%
This work has made used of CasJobs (original site http://casjobs.sdss.org/CasJobs). CasJobs was originally developed by the Johns Hopkins University/ Sloan Digital Sky Survey (JHU/SDSS) team. With their permission, MAST used version 3.5.16 to construct three CasJobs-based tools for GALEX, Kepler, and the HSC.
%
The authors would like to thank Rick White and Scott Fleming at Space Telescope Science Institute (and Jennifer Wiseman at NASA’s Goddard Space Flight Center for putting us in contact with one another) for their guidance in using the Hubble catalogue and CasJobs to find isolated point sources for use in the Monte Carlo analysis.
The authors would also like to thank 
Anthony Brown,
and Carme Jordi 
for their comments on earlier versions of this paper.
\end{acknowledgements}
\bibliographystyle{aa}
\bibliography{seapipe} 

\begin{appendix} 
\section{Finding isolated sources in the Hubble archive}
\label{appdx:hubble}

The Catalog Archive Server Jobs System (CasJobs) service which allows queries, phrased in the Structured Query Language (SQL), to be executed was used to extract candidate isolated point sources from the HSCv3. This service is available at: http://mastweb.stsci.edu/hcasjobs/ (original site http://casjobs.sdss.org/CasJobs).

The tables used were (note that in these descriptions match refers to a source in the HSC which has been found by cross-matching the sources found in the Hubble images):
\begin{itemize}
\item{{\it SumPropMagAper2Cat} which summarises the properties of each match based on sources with valid Source Extractor aperture magnitudes. This table contains the parameter {\it NumImages} which is the number of combined images (exposures with the same filter, same camera, and within the same visit) corresponding to the match.}
\item{{\it ClosestMatch} which contains the nearest neighbour for each match in table {\it SumPropMagAper2Cat}, provided there is a neighbour that lies
with $1^{\prime\prime}$  of the match.}
\item{{\it closeMatch} which contains, for each match,  a list of other matches that lie within $1^{\prime\prime}$  of its position.}
\item{{\it SumMagAper2} this table provides Source Extractor aperture magnitude information for each match. There is a row for every {\it Filter} and {\it Detector} combination which have observed the source corresponding to the {\it MatchID}.  Note that the table {\it SumPropMagAper2Cat} only contains summary information for the {\it MatchID} and does not contain magnitudes.}
\item{{\it DetailedCatalog} which contains the properties of each source in the Hubble Source Catalog; specifically of interest for this analysis is {\it Flags}, which is a bit encoded representation of source properties: 0 for point source, 1 for extended source, 4 for saturated source for every observation.}
\item{{\it XMatchV2} which contains, for each match,  a list of possible corresponding matches from HSCv2. }
\end{itemize}

The procedure for extracting the isolated candidates is as follows:
\begin{itemize}
\item{All {\it MatchID}s where {\it NumImages} $>3$ in the table {\it SumPropMagAper2Cat} and which are not present in the table {\it ClosestMatch} are extracted. The request for multiple images should filter out artefacts, while the requirement for the absence of the {\it MatchID} in the {\it ClosestMatch}  table should mean that there is no neighbouring source within $1^{\prime\prime}$.}
\item{The positions (RA,Dec) of the surviving {\it MatchID}s are extracted from the {\it SumPropMagAper2Cat} table.}
\item{The {\it Filter}, {\it Detector}, {\it MagMed} (median value of the contributing aperture magnitudes for the {\it Filter} and {\it Detector} combination) values for the {\it MatchID}s are extracted from the {\it SumMagAper2} table.}
\item{At this stage all rows with {\it MagMed} $>21.0$ are removed. This has the effect of removing candidate {\it MatchID}s if they are never observed in any {\it Filter} and {\it Detector} combination at a magnitude brighter than 21.}
\item{The {\it Flags} values from the {\it DetailedCatalog} are extracted for all the surviving {\it MatchID}s. Note that there is a {\it Flags} for every observation, so there are multiple {\it Flags} values per {\it MatchID}.}
\item{The {\it MatchID2} (the HSCv2 MatchID) and {\it DistArcSec} (angular separation of HSCv2 and HSCv3 matches in arc-seconds) values are extracted from the {\it XMatchV2} table for those surviving {\it MatchID}s which have an entry.}
\item{The extracted data is download from CasJobs in a CSV format.}
\end{itemize}
\subsection{Cross-matching Hubble data with Gaia}

The procedure cross-matching the isolated candidates is as follows:
\begin{itemize}
\item{Collate all the {\it Flags} for each Hubble {\it MatchID} and compute the average value.}
\item{Discard all Hubble {\it MatchID}s for which this average value is $\ge 0.1$ (keep only point sources).}
\item{Discard all Hubble {\it MatchID}s with a non-unique match to HSCv2.}
\item{Pick the most appropriate value to use as the Hubble magnitude (see Section~\ref{sec:hubble_filters}).}
\item{Cross-match the surviving candidates against the Science Alerts Database \citep{scienceAlerts}.}
\begin{itemize}
\item{Calculate the HP12 (healpix level 12) pixel corresponding to the candidate position to find the HP5 table in which the data may be stored, checking the neighbouring HP5 tables if the HP12 pixel is on the edge of an HP5 pixel.}
\item{Extract all the sourceIDs within $0.5^{\prime\prime}$  of the Hubble {\it MatchID}.}
\item{Discard if no Gaia data is found, or more than one SourceID is recovered.}
\end{itemize}
\item{For all Gaia SourceIDs which match to the Hubble {\it MatchID}s, extract their position, magnitude, astrometric parameters, and AEN (astrometric excess noise) from the DR2 archive.}
\item{Calculate the angular separation between the Hubble and Gaia positions and discard all matches with an angular separations $>0.1^{\prime\prime}$.}
\item{The Gaia track tables \citep{Torra21} are used to propagate the DR2 sourceIDs (used by IDT and the science alerts database) to DR3 and then to DR4 sourceIDs. At each stage only sources with one-to-one mappings of sourceIDs between cycles are kept, those with splits or merges are discarded.}
\end{itemize}

\subsection{Hubble filters}
\label{sec:hubble_filters}

For each Hubble {\it MatchID} there are multiple magnitude estimates, for different {\it Filter} and {\it Detector} combinations. We select the closest available filter to the Gaia pass-band, and use its recorded magnitude as the Hubble magnitude for the analysis presented here.

The priority ordering for which filter to use for the magnitude is as follows, first the WFC3 filters:
F350LP, F200LP, {\bf F606W}, F625W, {\bf F555W}, F475X, F600LP, F476W, {\bf F814W}, F775W, {\bf F850LP}, F438W, F621M, F689M, F763M, {\bf F547M}, F845M
then the WFC2 filters:
F165LP, F130LP, F569W, F622W, F675W, F702W, F791W, F785LP, F450W, F439W, F467M \\
Note that the filters in bold are also available for WFC2 
\end{appendix}

\end{document}